\documentclass[journal]{IEEEtranTIE}
\usepackage[linesnumbered,ruled,lined]{algorithm2e}
\usepackage{amssymb}

\usepackage{graphicx}
\usepackage{amsmath}
\usepackage{amssymb}
\usepackage{subfigure}
\usepackage{booktabs}
\usepackage{cite}
\usepackage{color}
\usepackage{balance}
\usepackage{multirow}
\usepackage{makecell}
\usepackage{makeidx}

\newtheorem{remark}{Remark}

\newcommand{\eg}{\textit{e}.\textit{g}.}

\begin{document}

\title{Experimental Validation of Multi-lane Formation Control for Connected and Automated Vehicles\\ in Multiple Scenarios}

\author{Mengchi Cai$^{1}$, Qing Xu*$^{1}$, Chunying Yang$^{1, 2}$, Jianghong Dong$^{1}$, Chaoyi Chen$^{1}$, Jiawei Wang$^{1}$, \\Jianqiang Wang$^{1}$, and Keqiang Li$^{1}$
\thanks{This work was supported by National Key Research and Development Program of China (2021YFB1600402), National Natural Science Foundation of China (52072212), National Natural Science Foundation of China, the Major Project (52131201), National Natural Science Foundation of China, the Major Project (61790561).}
\thanks{$^{1}$School of Vehicle and Mobility, Tsinghua University, Beijing 100084, P.~R.~China.}
\thanks{$^{2}$School of Transportation Science and Engineering, Beihang University, Beijing 100191, P.~R.~China.}
\thanks{$^{*}$Corresponding author: Qing Xu, E-mail: qingxu@tsinghua.edu.cn}
}

\maketitle

\begin{abstract}

Formation control methods of connected and automated vehicles have been proposed to smoothly switch the structure of vehicular formations in different scenarios. In the previous research, simulations are often conducted to verify the performance of formation control methods. This paper presents the experimental results of multi-lane formation control for connected and automated vehicles. The coordinated formation control framework and specific methods utilized for different scenarios are introduced. The details of experimental platform and vehicle control strategy is provided. Simulations and experiments are conducted in different scenarios, and the results indicate that the formation control method is applicable to multiple traffic scenarios and able to improve formation-structure-switching efficiency compared with benchmark methods.

\end{abstract}


\markboth{}%
{}


%
\section{Introduction}
\label{intro}
%

Multi-vehicle coordination methods are commonly utilized in multiple scenarios. Compared with single-vehicle automated control that considers objective of only the controlled vehicle, multi-vehicle coordination gathers information from vehicles and performs global optimization. Typical scenarios of multi-vehicle control include single-lane platooning~\cite{naus2010string,zheng2016distributed,wang2020controllability}, coordinated lane changing~\cite{wu2020emergency,luo2016dynamic,li2018consensus}, conflict resolution at ramps and bottlenecks~\cite{kato2002vehicle,ntousakis2016optimal,xu2020bi}, and scheduling at intersections~\cite{malikopoulos2018decentralized,bian2019cooperation,yu2019corridor}, etc. Existing research reveals that multi-vehicle coordination has great potential to guarantee driving safety, improve traffic efficiency, and reduce energy consumption~\cite{cai2021formationb,cai2021formationc,chen2021mixed}, compared with single-vehicle control. However, there is a lack of multi-vehicle coordination methods that consider both longitudinal and lateral behavior of vehicles in multi-lane scenarios.

Formation control is a classical problem in multi-agent systems, \eg\,robots and unmanned aerial vehicles (UAVs). Agents in a formation share information and cooperatively plan their motion to perform multiple maneuvers, \eg\,formation structure switching. Application scenarios of multi-agent formation control include coordinated flying of UAVs~\cite{dong2014time}, searching and exploring tasks of grounded robots~\cite{cheah2009region,macdonald2011multi}, and coordinated operation of underwater vehicles~\cite{li2016receding}, etc. Due to the similar nature of multi-agent systems and multi-vehicle network, it is natural for multi-vehicle control to learn from multi-agent formation control methods.

Formation control for CAVs is a specific case of multi-vehicle coordination in multi-lane scenarios. As an extension of single-lane platoon in multi-lane scenarios, formation control considers vehicles that are driving close to each other as a group, which is also called as a formation or a convoy, and cooperatively plan both of their longitudinal and lateral motion according to the scenarios and demands of vehicles. Existing research regarding on-road formation control for CAVs includes planning for geometric structure switching, formation maintaining control, etc. A vehicle-infrastructure cooperation method is proposed in~\cite{marinescu2012ramp} to organize vehicles to occupy moving slots, so as to perform ramp merging and leaving maneuvers. Communication topology and stable feedback controller are designed in~\cite{marjovi2015distributed} to enable vehicular formations to steadily maintain a desired formation structure and avoid collision. Coordinated lane changing behavior of vehicles are considered in~\cite{navarro2016distributed}, where vehicles are allowed to change lane in a formation. The idea of vehicle-to-target assignment and on-road formation control are combined in~\cite{cai2019multi,xu2021coordinated}, where vehicles are able to fully utilize lane capacity and improve traffic efficiency. Relative motion planning and conflict resolution methods in vehicular formations are proposed in~\cite{cai2021formationa,cai2021formationb} to avoid collision between vehicles during formation structure switching process. Vehicles' preference on lanes is considered in~\cite{cai2021formationc}, and a multi-vehicle relative motion planning method is accordingly proposed to control vehicles to change to their preferred lanes while avoiding collision. The idea of multi-lane formation control has also been extended to multi-lane intersections to cooperatively organize vehicles' longitudinal and lateral behavior~\cite{xu2021coordinated,cai2021multi}. Other research regarding on-road formation control include~\cite{zheng2021distance, cao2021platoon, firoozi2021formation}.

Among the previous research about formation control, simulations are often conducted to validate performance of those methods. Simulations on multi-lane road segments indicate that formation control of CAVs in multi-lane scenarios is able to improve traffic efficiency and reduce fuel consumption under high traffic volume~\cite{cai2019multi,cai2021formationa}. Application results in multi-lane ramps and intersections show the ability of the formation control method to be applied to multiple scenarios~\cite{cai2021formationb,cai2021formationc,cai2021multi}. Although existing simulations have already revealed plenty of advantages of formation control methods, there is a lack of experimental validation, which is crucial to harvest the aforementioned theoretical benefits of multi-vehicle formation control.

The main contribution of this paper is that it carries out simulations and experiments to validate the multi-lane formation control method, including functional validation and performance analysis. The results indicate that the formation control method is applicable to multiple traffic scenarios and able to improve formation-structure-switching efficiency compared with the benchmark method.

The rest of this paper is organized as follows. Section~\ref{coord} introduces the coordinated formation control framework. Section~\ref{method} provides the detailed formation control methods in different scenarios. Section~\ref{simexp} carries out simulations and experiments to validate the formation control method. Section~\ref{conc} presents the conclusion of this paper.

%
\section{Formation Control Framework}
\label{coord}
%

\begin{figure}
\begin{center}
    \includegraphics[width=0.95\linewidth]{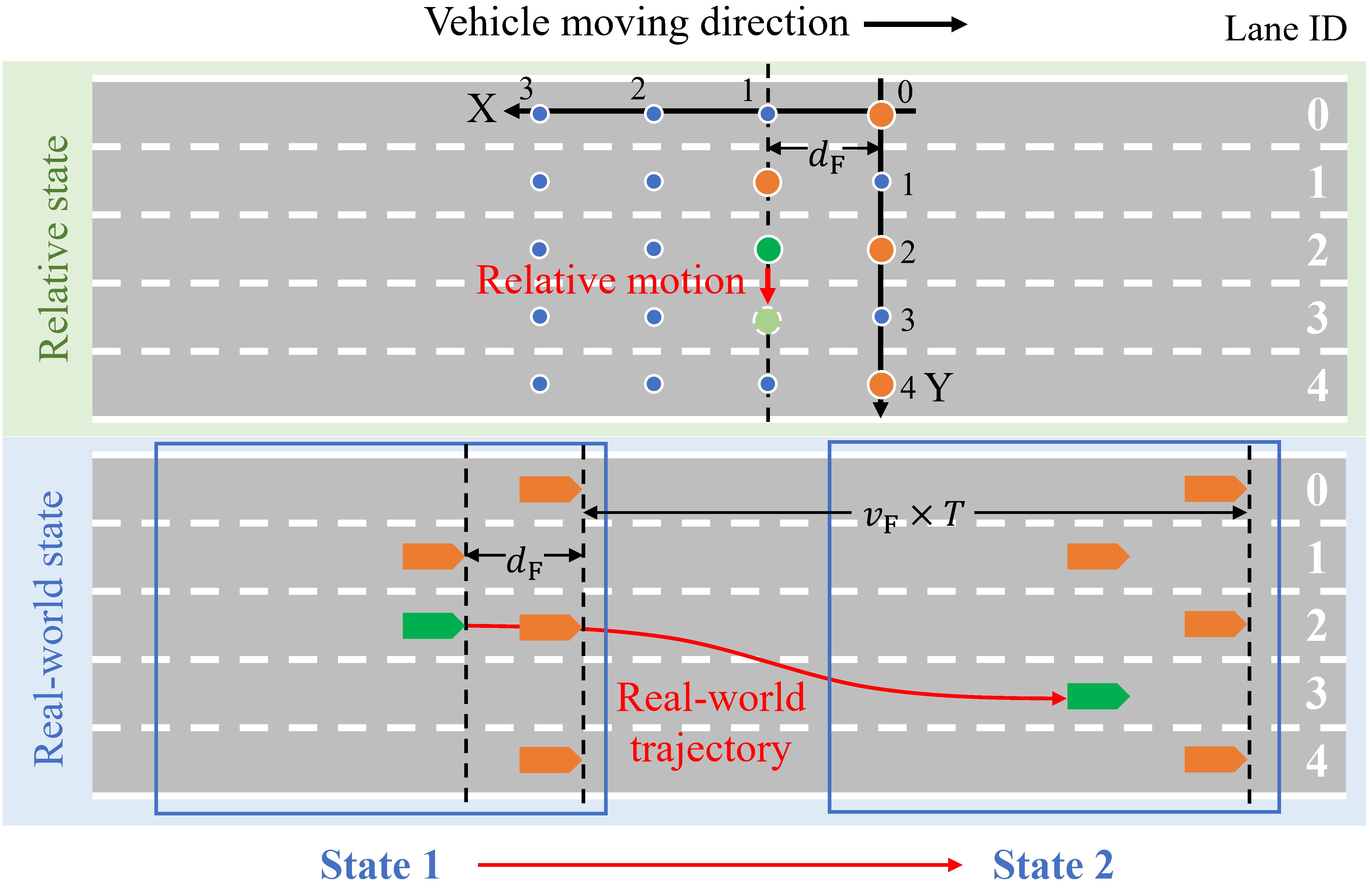}
    \caption{The relative coordinate system of vehicular formations and projection relationship between relative state and real-world state. }
    \label{bilevel}
\end{center}
\end{figure}

\begin{figure}
\begin{center}
    \includegraphics[width=0.95\linewidth]{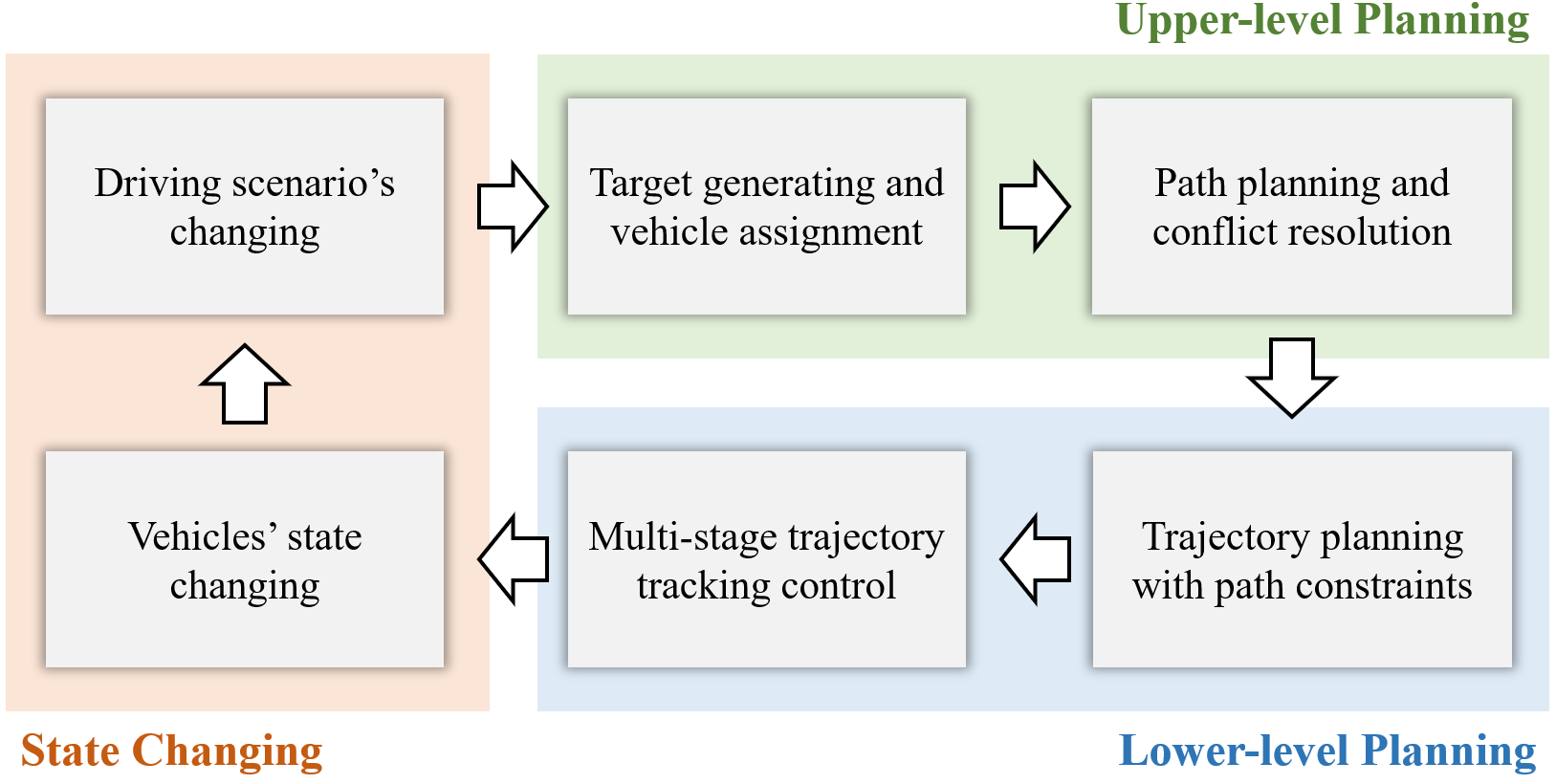}
    \caption{The coordinated formation control framework. }
    \label{process}
\end{center}
\end{figure}

Collision-avoidance of multiple vehicles is hard because vehicles have different driving expectations, and arbitrarily designed traffic rules may result in congestion and limit traffic efficiency. This study proposes a bi-level formation control framework to solve this problem. Some preliminary results of this study have been presented in earlier publications~\cite{cai2021formationa, cai2021formationb, cai2021formationc}.

In the upper level, motion of vehicles is planned in relative coordinate system (RCS), which describes the relative relationship between vehicles and moves together with the formation in a grid map, as shown in Fig.~\ref{bilevel}. In RCS, the position is discretized in both lateral and longitudinal direction. The discretization distance $d_\text{F}$ is safe enough for both single-lane following and cut-in movement from the adjacent lane. Vehicles move from one relative point to another in RCS, and coordinated path planning methods on grid maps are utilized to plan paths for vehicles and avoid collision. The time that vehicles move from one relative point to another is set as the same time cycle $T_\text{F}$. During one time cycle, a vehicle is allowed to move to the other four points, which correspond to longitudinal position adjustment and lateral lane change respectively, or stay at its current position. The movement of vehicles in RCS is synchronous, so as to clearly organize cooperative behavior and avoid collision.

In the lower level, the series of relative points that vehicles will pass through are used to plan real-world trajectories for vehicles in the geodetic coordinate system (GCS). Vehicles plan their real-world trajectories to travel through these projected points and perform trajectory tracking with spatiotemporal constraints, including geometric constraints on curves and tracking time constraints. An example of a vehicle (marked in green) changing lane to form five-vehicle formation with the other four vehicles is shown in Fig.~\ref{bilevel}.

The working process of the formation control framework is shown in Fig.~\ref{process}. The process starts with the changing of driving scenarios, including changing of lane number, lane topology, vehicles' driving goals, etc. Then, four steps are taken to switch the structure of formations to adapt to the changing scenarios:
\begin{enumerate}
\item \textbf{Target generating and vehicle assignment}. Target points are generated according to the number of lanes demands of vehicles. If vehicles have preference on lanes, targets are generated accordingly on the specific lanes, and if not, targets are generated evenly on all the lanes. Vehicles are then assigned to different targets to minimize the assigning cost, \eg\, the maximum traveling time or the total covered distance. 
\item \textbf{Path planning and conflict resolution}. Paths are planned for vehicles to  travel to their assigned targets in the grid RCS. The paths of vehicles may cross or overlap, which may result in collision. Conflict-resolution methods are utilized to adjust or replan paths of vehicles to clear those collision-leading conflicts.
\item \textbf{Trajectory planning with path constraints}. Vehicles should pass through those road points that are projected from the path points in RCS. Smooth trajectories are planned the connect those road points, \eg\, B$\acute{\text{e}}$zier curves.
\item \textbf{Multi-stage trajectory tracking control}. Vehicles should arrive at the projected road points with time constraints, in order to follow the motion steps in RCS and avoid collision. Since the trajectory of a vehicle consists of multiple curve segments, multi-stage tracking methods are utilized and control inputs are recalculated one the following of one segment is finished, in order to resolve accumulated tracking error.
\end{enumerate}

The above four steps change the state of vehicles to adapt to the changing scenarios, and if the driving scenario changes again, the process will start again. The upper-level planning parts can be conducted in a centralized planner, such as the roadside units and the cloud computation platform, which needs to gather information from vehicles and road. The lower-level planning parts can be conducted in both centralized or decentralized way, since the planning of each vehicle is independent.

\begin{remark}
The proposed formation control framework is applicable for not only fully-CAV environment, but also for a mixture of both CAVs and human driven vehicles (HDVs) and fully-HDV environment. HDVs are treated differently according to their connectivity. If an HDV is able to receive control instructions from the centralized planner, the trajectory following process can be conducted by the driver, since the instructions are easy to understand and conduct, like ``changing to the right lane while keeping the current speed'' and ``keeping a given distance from the preceding vehicle''. If an HDV is not able to receive instructions, it will be treated as obstacle that blocks one lane, and the other vehicles will change the structure of the formation to drive on the remained lanes.
\end{remark}

%
\section{Methodologies}
\label{method}
%

In this section, the methods that are used for multi-vehicle formation control are introduced in detail, and some examples are provided to show the demonstration of formation control on multi-lane roads.

%
\subsection{Target Generating and Vehicle Assignment}
\label{targetgenerating}
%

\begin{figure}
\begin{center}
    \subfigure[Parallel structure]{
    \includegraphics[width=0.45\linewidth]{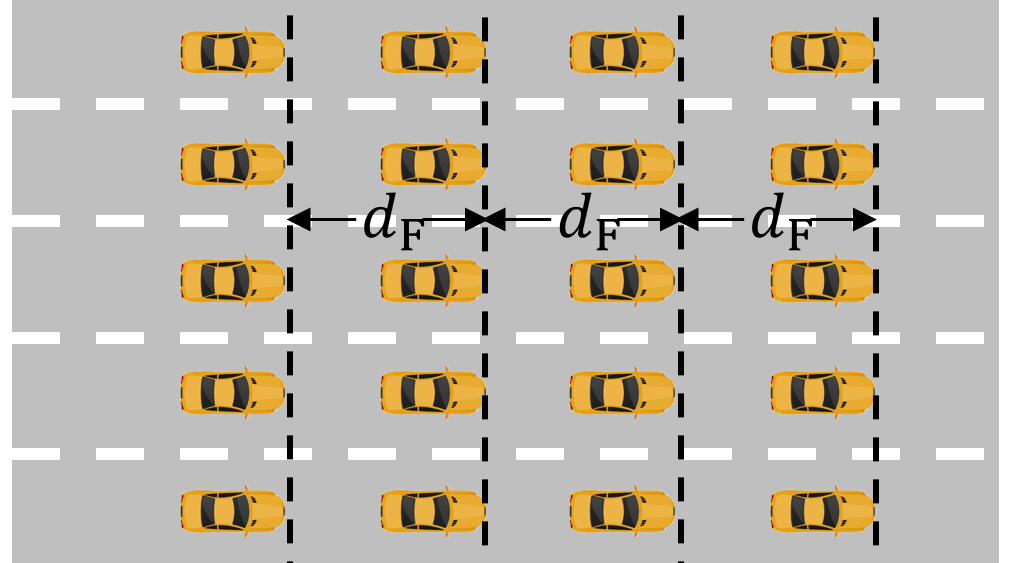}
    \label{formation1}}
    \subfigure[Interlaced structure]{
    \includegraphics[width=0.45\linewidth]{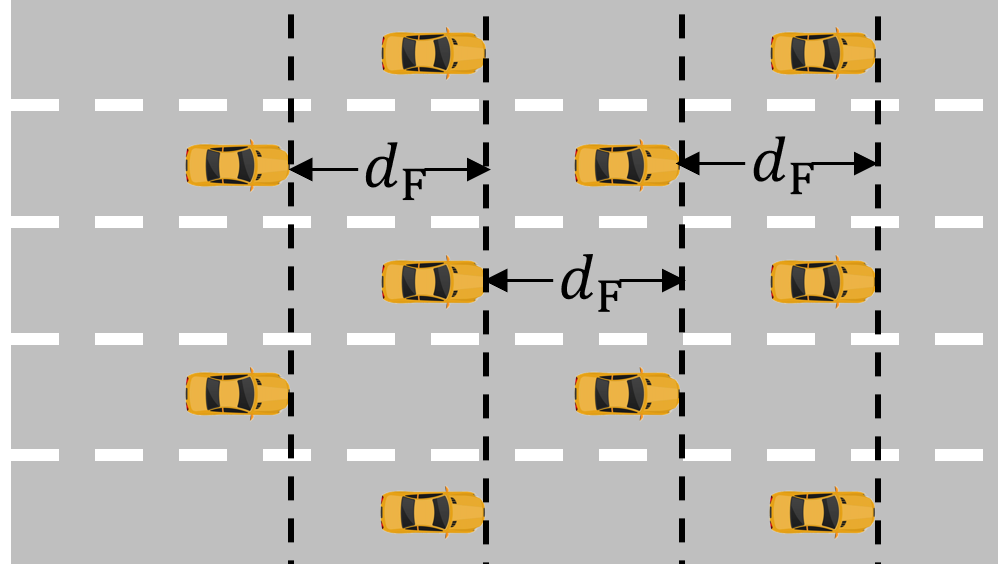}
    \label{formation1}}
    \caption{Common formation geometric structures. }
    \label{structure}
\end{center}
\end{figure}

Similar to single-lane platooning where vehicles drive together in a desired geometric structure (following distance), vehicles drive as formations with specific geometric topology on multi-lane roads. Typical formation structures on multi-lane roads include interlaced structure, parallel structure, etc. Existing research reveals that although the parallel structure has higher vehicle density, the interlaced structure is more suitable for multi-lane vehicle coordination considering lane-changing efficiency and driving safety~\cite{marjovi2015distributed, cai2019multi}. Thus, the interlaced structure is chosen as the standard driving structure of vehicular formation on multi-lane road segments. The occupation of vehicles on relative points in these two structures are shown in Fig.~\ref{structure}.

Total number of vehicles is equal to the number of vehicles in a formation. Besides, vehicles may have preference on lanes, \eg\, when vehicles are approaching an intersection and have to change to specific lanes according to their routes in the intersection. Thus, the number of targets on each lane should be equal to the number of vehicles with the according lane preference.

After generating the targets in RCS, vehicles should be assigned to specific targets to form a one-to-one matching relationship. An optimal assignment is built by minimizing total cost for all the vehicles to travel to their targets. The cost can be defined in various ways, \eg\, the travelling time, the covered distance, etc. After determining the assignment cost between each vehicle and target, the cost matrix $\mathcal{C}$, whose element on the $i$-th row and the $j$-th column represents the cost to assign vehicle $i$ to target $j$, is defined as:
\begin{eqnarray}
\mathcal{C}=[c_{i,j}]\in \mathbb{R}^{N\times N},\ i,j\in \mathbb{N}^+,
\end{eqnarray}
where $N$ is the total number of vehicles.

The assignment matrix $\mathcal{A}$, whose element on the $i$-th row and the $j$-th column represents whether vehicle $i$ is assigned to target $j$, is defined as:
\begin{eqnarray}
&\mathcal{A}=[a_{i,j}]\in \mathbb{R}^{N\times N},\ i,j\in \mathbb{N}^+, \ \\
&a_{i,j}=
\begin{cases}
1, \ \text{if vehicle $i$ is assigned to target $j$},\notag \\
0, \ \text{otherwise}.
\end{cases}
\end{eqnarray}

Vehicles' preference on lanes is accordingly transformed to the preference on targets. The preference matrix $\mathcal{P}$ is defined to describe the preference of vehicles on different targets. The element on the $i$-th row and the $j$-th column of $\mathcal{P}$ represents whether vehicle $i$ prefers target $j$, and $\mathcal{P}$ is defined as:
\begin{eqnarray}
&\mathcal{P}=[p_{i,j}]\in \mathbb{R}^{N\times N},\ i,j\in \mathbb{N}^+, \ \\
&p_{i,j}=
\begin{cases}
1, \ \text{if vehicle $i$ has preference on target $j$},\notag \\
M, \ \text{otherwise},
\end{cases}
\end{eqnarray}
where $M$ is a positive number that is large enough to prevent a vehicle from being assigned to a target. Then, the assignment problem with target preference can be modelled as:
\begin{alignat}{2}
\min\quad & \sum_{i=1}^N\sum_{j=1}^N (c_{i,j}\times p_{i,j}\times a_{i,j}),\label{eqn - lp2}\\
\mbox{s.t.}\quad
&\sum_{i=1}^N a_{i,j}=\sum_{j=1}^N a_{i,j}=1,\notag \\
&i,j\in \mathbb{N}^+.\notag
\end{alignat}
where $[c_{i,j}]$ and $[p_{i,j}]$ are the given cost matrix and preference matrix, and $[a_{i,j}]$ is the variable assignment matrix. The assignment problem can be solved using Hungarian algorithm~\cite{19kuhn1955hungarian}, the simplex algorithm, etc.

%
\subsection{Path Planning and Conflict Resolution}
\label{pathplanning}
%

\begin{figure*}
\begin{center}
    \subfigure[Steps of formation structure switching]{
    \includegraphics[width=0.95\linewidth]{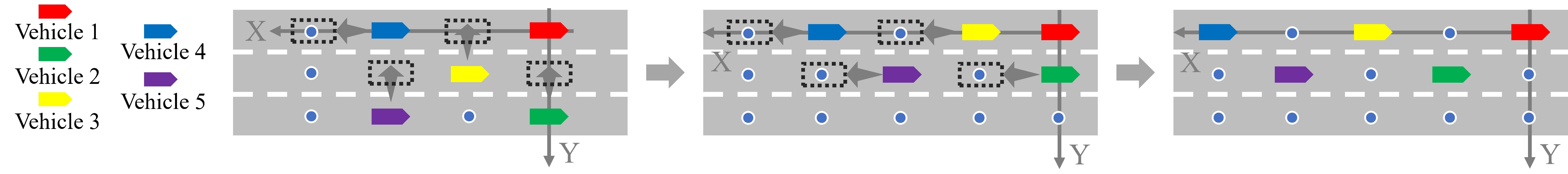}
    \label{steps1}}
    \subfigure[Real-world trajectories of vehicles]{
    \includegraphics[width=0.95\linewidth]{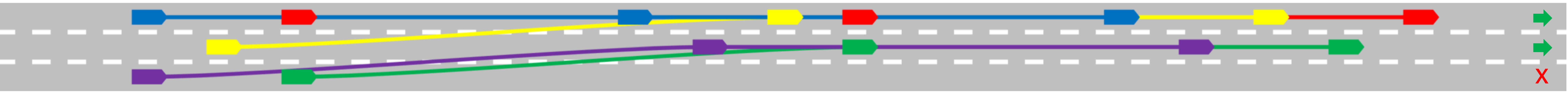}
    \label{traj1}}
    \caption{Example of multi-vehicle formation control without lane preference.}
    \label{example1}
\end{center}
\end{figure*}
\begin{figure*}
\begin{center}
    \subfigure[Steps of formation structure switching]{
    \includegraphics[width=0.95\linewidth]{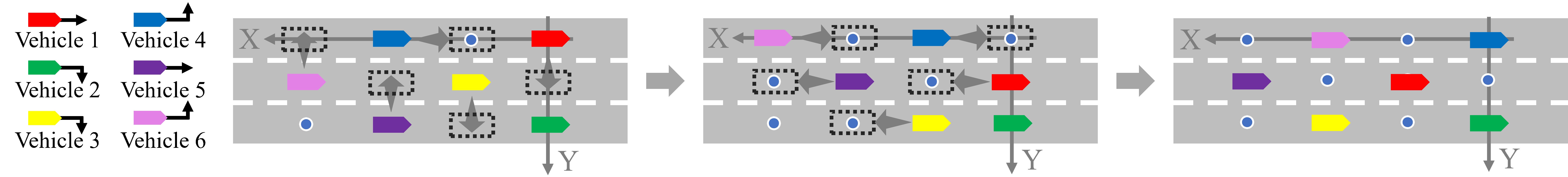}
    \label{steps2}}
    \subfigure[Real-world trajectories of vehicles]{
    \includegraphics[width=0.95\linewidth]{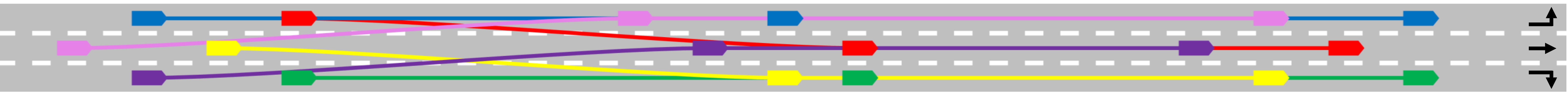}
    \label{traj2}}
    \caption{Example of multi-vehicle formation control with lane preference.}
    \label{example2}
\end{center}
\end{figure*}

After building the assignment between vehicles and targets, the paths for vehicles to travel to targets are then planned. Paths are defined differently from trajectories in this paper. Paths connect relative points in RCS and describe vehicles' motion in a coarse-grained manner, while trajectories present the movement of vehicles in GCS. Traditional single-vehicle path planning method on grid maps such as A* can be utilized for single-vehicle path planning~\cite{hart1968formal}. Possible conflicts may exist when vehicles arrive at the same point at the same time, or when their paths cross or overlap during the same time cycle $T_\text{F}$.

The way to resolve conflicts between vehicles depends on whether vehicles have preference on targets. If vehicles don't have preference on targets, these conflicts can be resolved by letting one vehicle waiting at its current relative point to let the other go at first, or switching the targets of two vehicles. If vehicles have preference on targets, simply switching targets of two vehicles may not be a feasible way to resolve conflicts. There are some path planning methods in the field of multi-agent pathfinding (MAPF) that can be utilized to solve this problem, \eg\, conflict-based searching (CBS)~\cite{sharon2015conflict}, multi-agent A*~\cite{wagner2011m}, etc. The CBS method searches global optimal solution in a conflict tree, and its property guarantees that it can return at least one solution, if there exists one. The A* family plans paths for vehicles based on defined priority, and sometimes results in no solution, even if there exists one. Thus, CBS is chosen as the multi-vehicle path planning method when vehicles have preference on lanes in this study. The time complexity and accelerating methods of CBS are the focuses of recent research~\cite{felner2018adding, li2019improved}. Given that CBS searches solutions in an expanding tree, and iteratively replaces the current solution with a better one, one way to guarantee completeness of CBS is to set a time bound. If the global optimal solution is found within the bound, then it is returned, and if the time bound is reached, the current local optimal solution is returned. If the time bound is reached and no feasible solution is achieved, the algorithm returns no solution. More details about the conflict resolution and proof of properties can be found in~\cite{cai2021formationb} and \cite{cai2021formationc}.

%
\subsection{Trajectory Planning and Tracking Control}
\label{trajectoryplanning}
%

The output of path planning is a series of relative points in RCS. Those points are projected to GCS and the projected real-world road points are used as inputs for vehicular trajectory planning. There are many types of curves that can be chosen to generate trajectories for vehicles, and B$\acute{\text{e}}$zier curve is one of the most commonly used one~\cite{gonzalez2015review}. Since there are possibly more than two road points for vehicles to pass through, the whole trajectory consists of several B$\acute{\text{e}}$zier curves and vehicles will perform multi-stage motion control to track the trajectory. In this paper, the cubic B$\acute{\text{e}}$zier curve with four control points is chosen for single-segment trajectory planning. 

Similar to most of the research regarding trajectory tracking control, this paper decouples the lateral and longitudinal control of vehicles. As for lateral control, a PID-based preview controller is designed to calculate steering angle of front wheels. As for longitudinal control, the coordinated motion of vehicles is divided into stages according to the synchronous moves in RCS, and the next stage begins when all the vehicles have finished the current stage. It is important to notice that the time for vehicles to finish tracking of one B$\acute{\text{e}}$zier-curve segment may be different, and the early arrived vehicles will keep the desired speed of formation and wait for the other vehicles. Thus, there may be some straight lines connected those B$\acute{\text{e}}$zier-curve segments. Optimal control method is a typical way to solve a tracking control problem with fixed time constraints. However, if the time $T_\text{F}$ is not known or restricted, a linear feedback controller is also a qualified candidate for longitudinal control. For more details about formation longitudinal control, please refer to~\cite{cai2021formationb}.

%
\subsection{Examples}
\label{examples}
%

Examples of multi-vehicle formation control are shown in Fig.~\ref{example1} and Fig.~\ref{example2}. In Fig.~\ref{example1}, five vehicles start from a three-lane interlaced structure and need to switch to a two-lane structure because the third lane becomes undrivable. The steps of formation structure switching are shown in Fig.~\ref{steps1} and the real-world trajectories are presented in Fig.~\ref{traj1}. During the first step, vehicle 2, vehicle 3 and vehicle 5 change to the left lane, and vehicle 4 changes its speed to adjust longitudinal relative position. During the second step, all the vehicles but vehicle 1 adjust their longitudinal relative position and a five-vehicle two-lane interlaced formation structure is formed. In Fig.~\ref{example2}, six vehicles start from a three-lane interlaced structure and need to change to their preferred lanes according to their routes. The routes of each vehicle and steps of formation structure switching are shown in Fig.~\ref{steps2} and the real-world trajectories are presented in Fig.~\ref{traj2}. During the first step, vehicle 1, vehicle 3, vehicle 5 and vehicle 6 change to their preferred lane, and vehicle 4 changes its speed to adjust longitudinal relative position. During the second step, all the vehicles but vehicle 2 adjust their longitudinal relative position and a six-vehicle formation structure where all the vehicles are on their preferred lanes is formed.

%
\section{Simulations and Experiments}
\label{simexp}
%

In this section, simulations and experiments are carried out to validate the function of the proposed method and compare its performance with benchmark methods. 

%
\subsection{Performance Analysis Simulations}
\label{simu}
%

\begin{figure}
\begin{center}
    \includegraphics[width=0.95\linewidth]{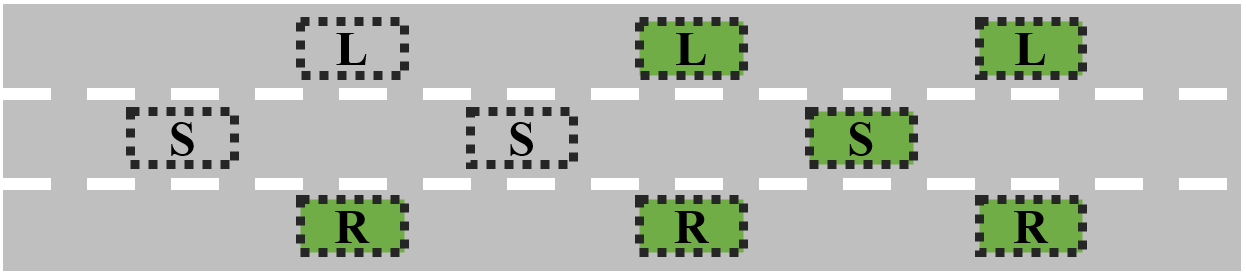}
    \caption{An example of target distribution in the simulations. ``L'', ``S'', and ``R'' represent that the targets should be occupied by vehicles that are turning left, going straight, and turning right respectively. The rectangles in green represent the targets that are matched with vehicles, and the hollow ones are not. This figure presents an example of the target distribution of six vehicles.}
    \label{targetexample}
\end{center}
\end{figure}

In order to evaluate the performance of the proposed multi-vehicle coordinated path planning algorithm, this study conducts simulations in the scenario with three lanes where vehicles start from a standard interlaced structure and have to change to their desired lanes. The simulations are conducted with different number of vehicles with lane preference. Targets are generated also according to the interlaced structure, and an example is presented in Fig.~\ref{targetexample}. The multi-agent priority-based A* algorithm is chosen as the benchmark method. The vehicle that is more forward in the formation is set with higher priority. Vehicles perform single-vehicle path planning according to the priority, and the paths of the planned vehicles are considered as obstacles for the latter vehicles. In order to compare performance fairy, both A* and CBS are utilized within RCS, and the starting state, goal state, and moving rules are all set as the same. The width of the grid map is set to 3, allowing vehicles to drive on the three lanes. The length of the grid map is set for all vehicles to move within the largest range limited by the position of targets. 

The success rate of the algorithm, maximum steps and total steps that vehicles take to change the structure of formation are chosen as indexes to evaluate the performance. Given that the proposed method consumes more time when the number of vehicles becomes bigger, the maximum running time of CBS is set to 2 seconds and 10 seconds respectively. Results of simulation are presented in Table~\ref{simutable1}, where total number of cases under each number of vehicles, failed times, success rate, maximum steps and total steps that vehicles take to change the structure of formation, and time consumption of the three methods are provided. The results indicate that the priority-based A* algorithm uses the minimum time, but results in high failed cases. In contrast, the CBS methods significantly improve success rate and get better performance in reducing steps that vehicles should take. The comparison between the two CBS methods with different time bound indicates that the CBS method is able to handle the hard cases where A* fails if given more computational time. Note that the failed times of five vehicles is larger than that of six vehicles due to the size of the grid map. According to the results of simulations, the number of vehicles is set to five or six in the following experiments.

\begin{table}[htbp]
\centering
\caption{Results of the simulations}
\label{simutable1}
\begin{tabular}{m{1em}m{2em}m{3.6em}m{2em}m{3em}m{2em}m{2em}m{2em}}
\toprule
Veh. No.& Total No. & Methods & Failed No. & Success rate & Max. steps & Total steps & Time \,(s)\\
\midrule
\multirow{4}*{5} & \multirow{4}*{243} &A*& 18 & 92.59\% & 3.86 & 12.34 & 0.01\\
\cmidrule{3-8}
&& CBS\,(2\,s)  & 2 & 99.18\% & 3.66 & 10.72 & 0.02\\
\cmidrule{3-8}
&& CBS\,(10\,s)  & 2 & 99.18\% & 3.66 & 10.72 & 0.02\\
\midrule
\multirow{4}*{6} & \multirow{4}*{729} &A*& 8 & 98.90\% & 4.08 & 15.36 & 0.01 \\
\cmidrule{3-8}
&& CBS\,(2\,s)  & 1 & 99.86\% & 3.78 & 13.16 & 0.04\\
\cmidrule{3-8}
&& CBS\,(10\,s)  & 0 & 100.00\% & 3.78 & 13.16 & 0.05\\
\midrule
\multirow{4}*{7} & \multirow{4}*{2187} &A*& 58 & 97.35\% & 4.49 & 18.86 & 0.01  \\
\cmidrule{3-8}
&& CBS\,(2\,s)  & 15 & 99.31\% & 4.06 & 16.10 & 0.09\\
\cmidrule{3-8}
&& CBS\,(10\,s)  & 1 & 99.95\% & 4.06& 16.10 & 0.09\\
\midrule
\multirow{4}*{8} & \multirow{4}*{6561} &A*& 270 & 95.88\% & 4.95 & 22.61 & 0.01  \\
\cmidrule{3-8}
&& CBS\,(2\,s)  & 307 & 95.32\% & 4.42 & 19.28 & 0.16\\
\cmidrule{3-8}
&& CBS\,(10\,s)  & 80 & 98.78\% & 4.42 & 19.25 & 0.16\\
\bottomrule
\end{tabular}
\end{table}

%
\subsection{Experimental Platform}
\label{plat}
%

\begin{figure}
\begin{center}
    \subfigure[Experimental platform]{
    \includegraphics[width=0.95\linewidth]{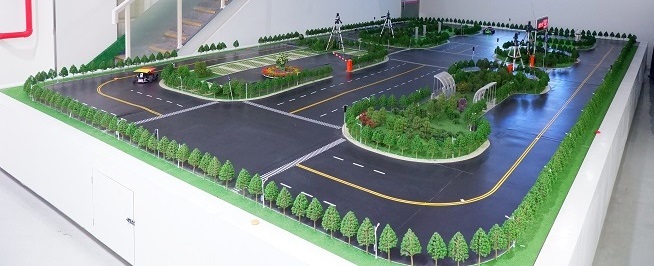}
    \label{city}}
    \subfigure[Experimental vehicles]{
    \includegraphics[width=0.95\linewidth]{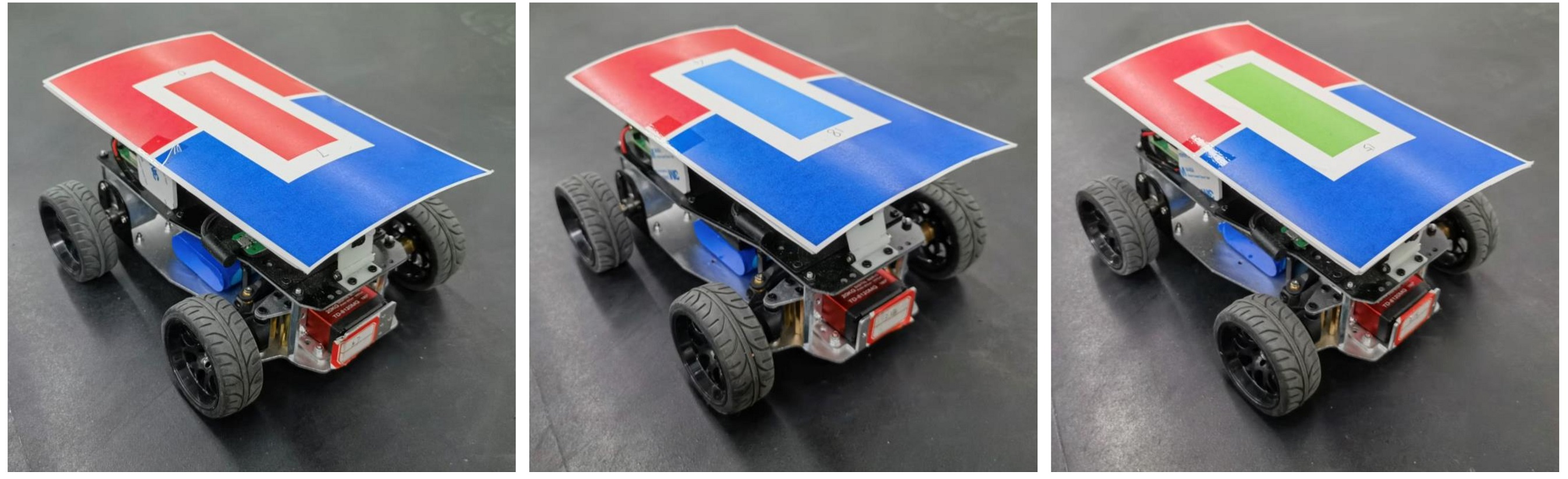}
    \label{expvehicle}}
    \caption{The experimental platform and vehicles. }
    \label{platformfig}
\end{center}
\end{figure}

The proposed multi-lane formation control method is validated on the connected micro-vehicle experimental platform built by Tsinghua University. The platform consists of multiple traffic scenarios, including multi-lane road segments and intersections, and is able to support experiments of single-vehicle automated control, multi-vehicle coordinated control, and human-machine cooperation, etc. Figures of the platform and experimental vehicles are presented in Fig.~\ref{platformfig}. The boards with different color on the top of vehicles are used to locate and identify vehicles driving on the platform by cameras. More details of the designing and function of the platform have been introduced in \cite{yang2021multi}.

\begin{figure}
\begin{center}
    \subfigure[Original lane utilization]{
    \includegraphics[width=0.46\linewidth]{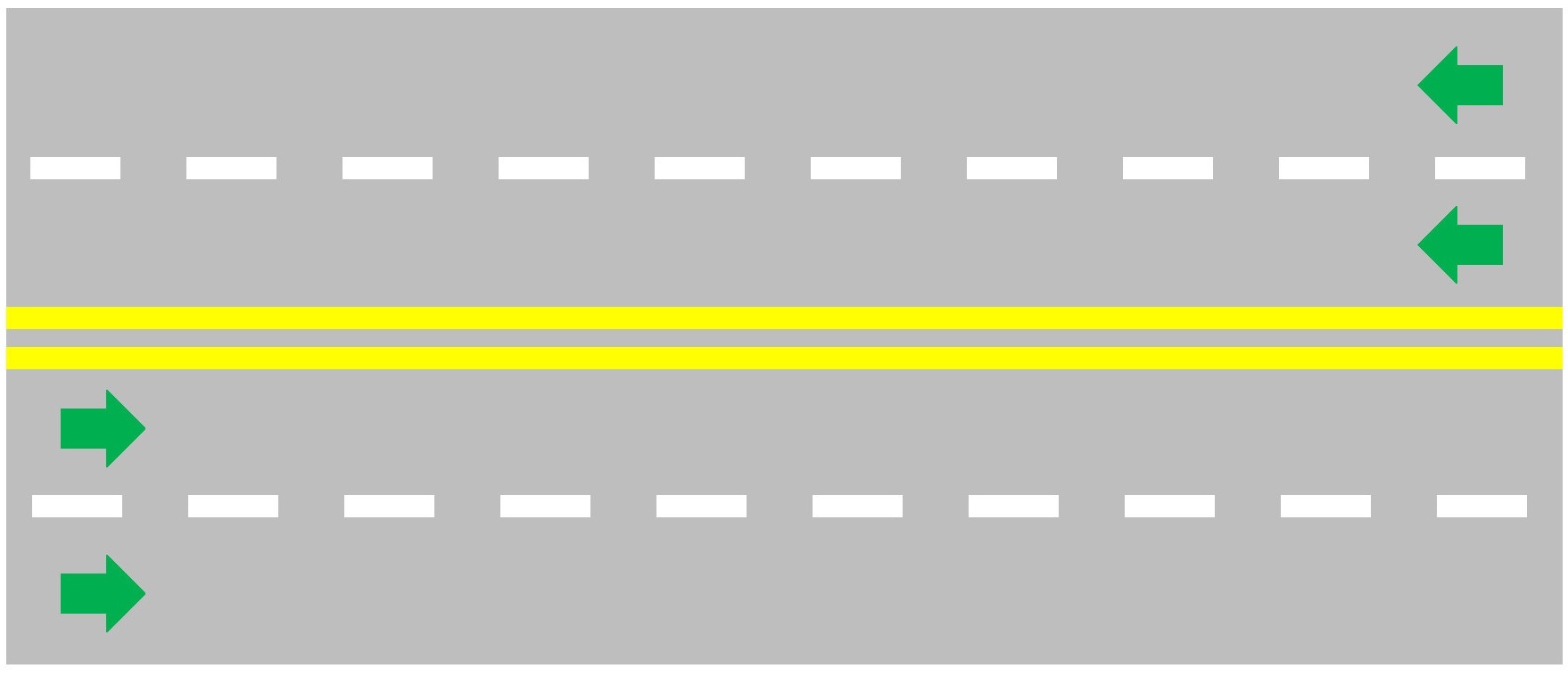}
    \label{formation1}}
    \subfigure[Adjusted lane utilization]{
    \includegraphics[width=0.46\linewidth]{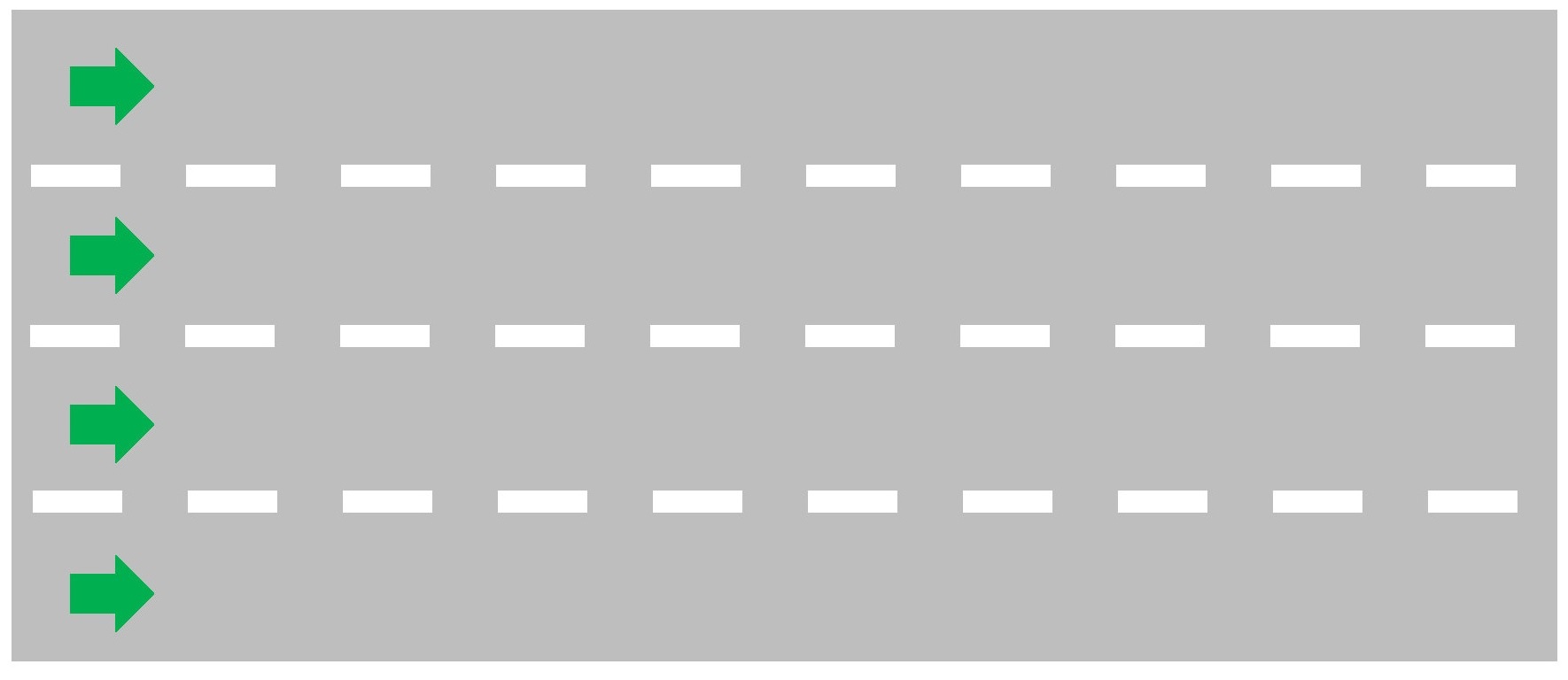}
    \label{formation1}}
    \caption{Lane utilization of the platform. }
    \label{laneutilization}
\end{center}
\end{figure}

The experiments in this paper are carried out on the central four-lane road segment. The chosen segment contains four lanes, where two lanes are used for each direction of traffic. In this paper, in order to validate the proposed multi-lane formation control method, multiple lanes are needed for vehicles to form desired structures. Thus, the original lane utilization is adjusted to a one-direction four-lane case, as shown in Fig.~\ref{laneutilization}. As for the adjusted lane utilization, vehicles drive from left to right on all the four lanes, and are allowed to change lane between every two adjacent lanes, which means that sometimes vehicles may cross the double-yellow solid lines.

A computer with CPU Intel CORE i7-8700@3.2GHz and 16G RAM serves as the centralized planner to gather information from the platform and vehicles, and calculate control inputs, including desired speed and steering angle, and send them back to vehicles accordingly. Actuators then control the motors to drive vehicles to follow the instructions. The end-to-end time delay of the system is around $100\,\mathrm{ms}$ and is omitted in the control process.

%
\subsection{Experimental Validation in Multiple Scenarios}
\label{vali}
%

\begin{figure}
\begin{center}
    \subfigure[Formation switching from one-lane structure to three-lane structure]{
    \includegraphics[width=0.99\linewidth]{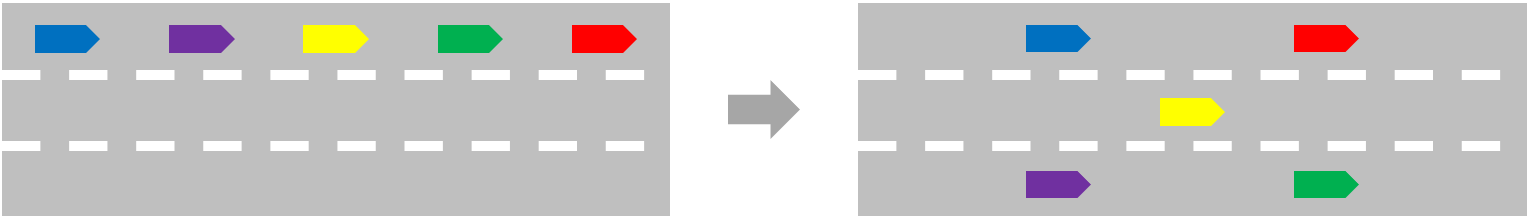}
    \label{formation1}}
    \subfigure[Formation switching from three-lane structure to two-lane structure]{
    \includegraphics[width=0.99\linewidth]{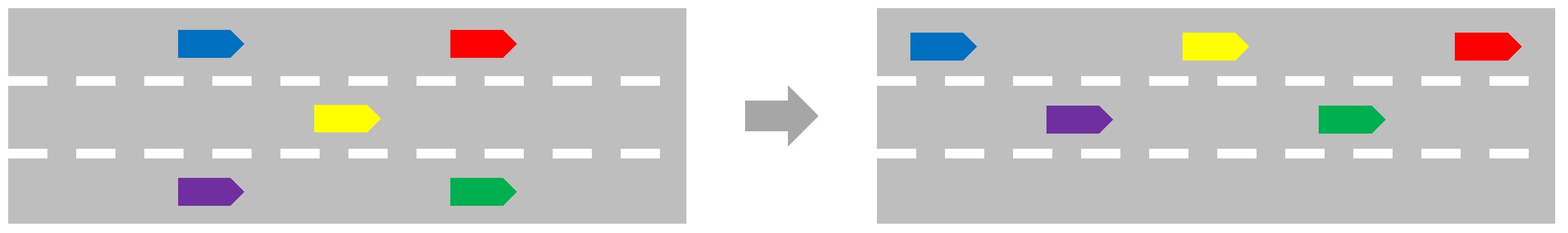}
    \label{formation1}}
    \subfigure[Formation switching from two-lane structure to one-lane structure]{
    \includegraphics[width=0.99\linewidth]{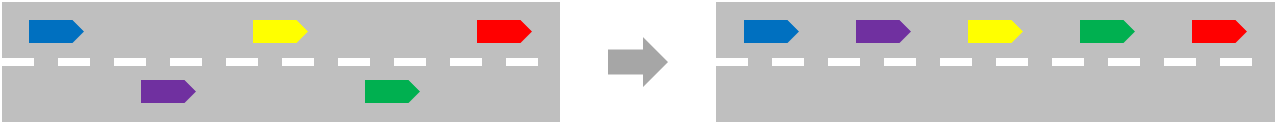}
    \label{formation1}}
    \subfigure[Formation switching at on-ramp merging area]{
    \includegraphics[width=0.99\linewidth]{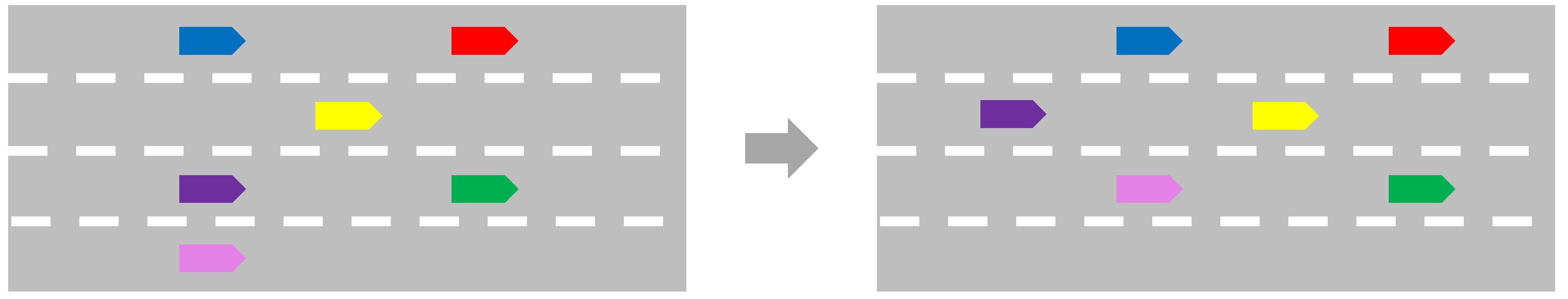}
    \label{formation1}}
    \subfigure[Formation switching at off-ramp leaving area]{
    \includegraphics[width=0.99\linewidth]{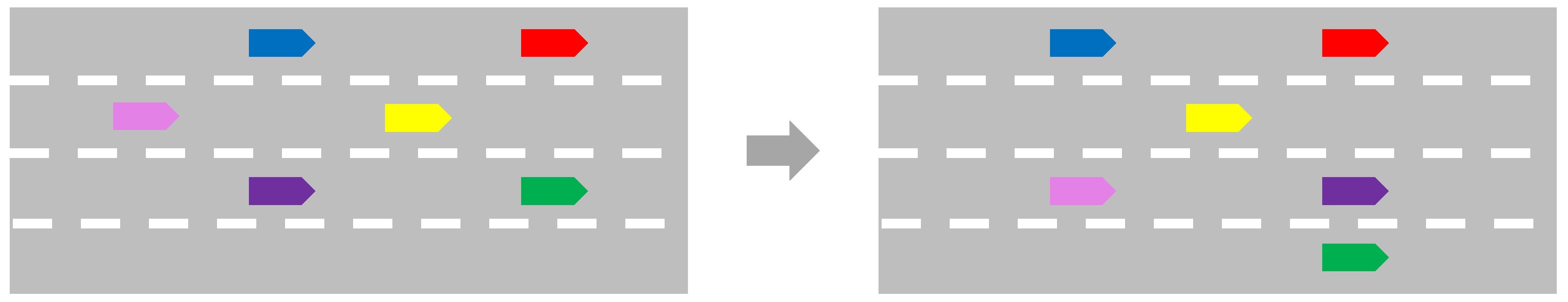}
    \label{formation1}}
    \subfigure[Formation switching in emergency leaving scenario]{
    \includegraphics[width=0.99\linewidth]{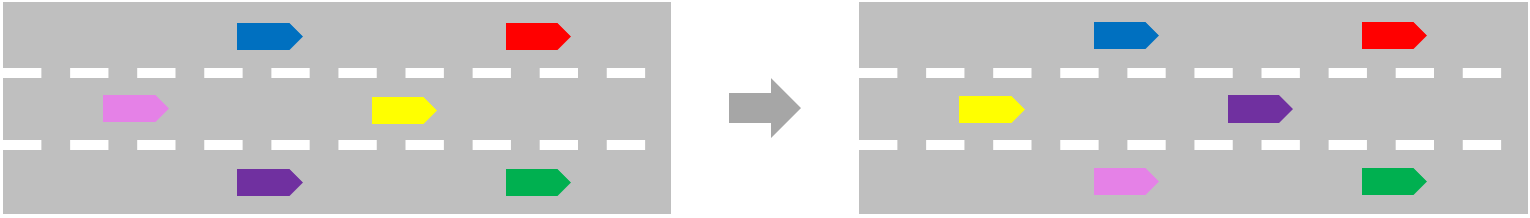}
    \label{formation1}}
    \subfigure[Formation switching in cooperative lane changing scenario]{
    \includegraphics[width=0.99\linewidth]{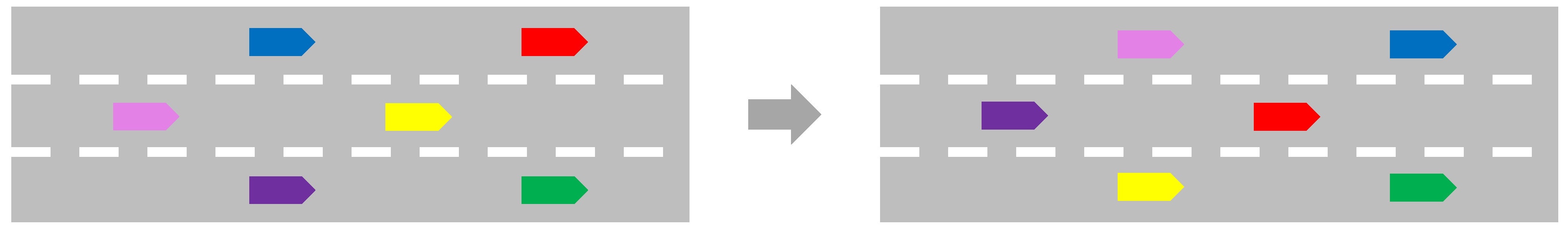}
    \label{formation1}}
    \caption{Scenarios of functional validation experiments. }
    \label{scenarios}
\end{center}
\end{figure}

\begin{figure*}
\begin{center}
    \subfigure[Formation switching from one-lane structure to three-lane structure]{
    \includegraphics[width=0.95\linewidth]{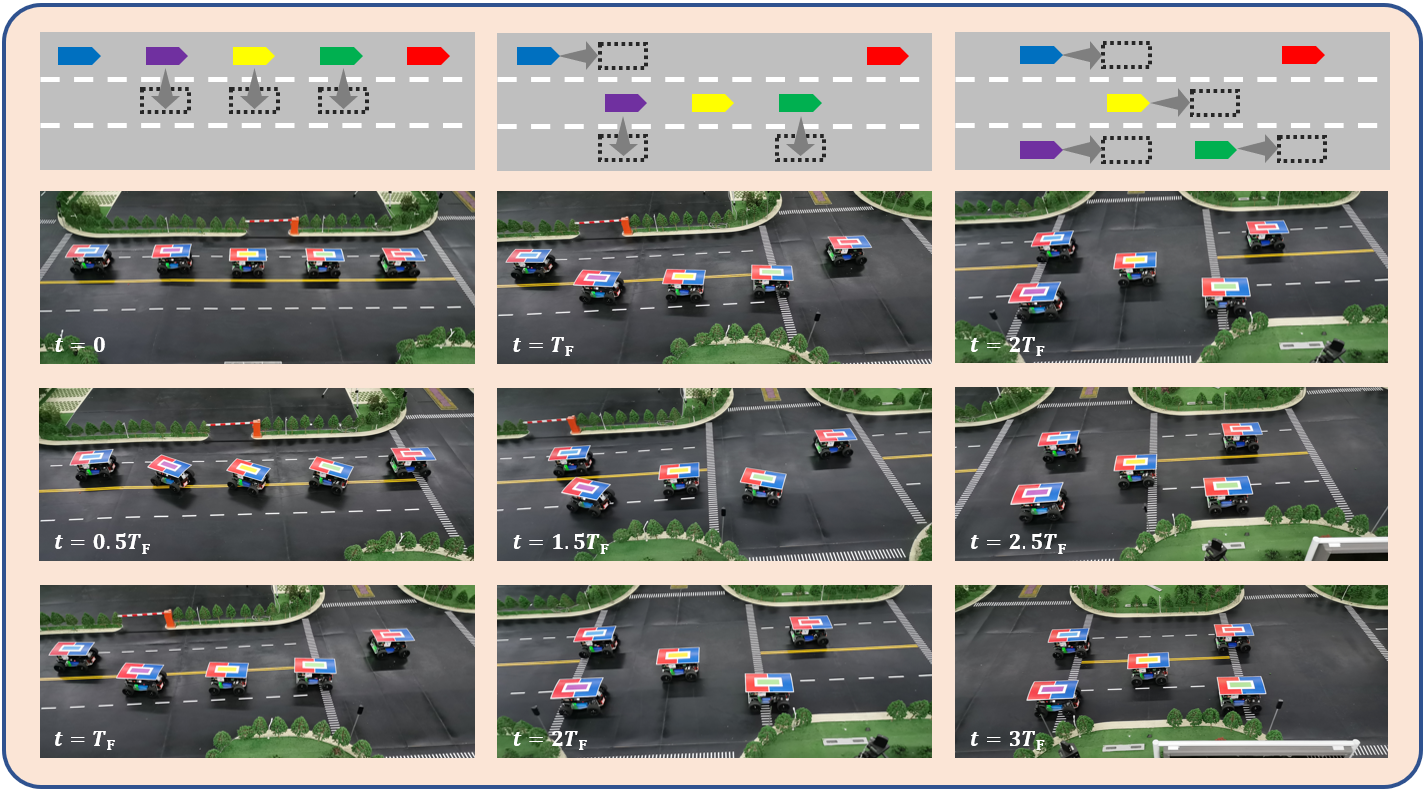}
    \label{formation1}}
    \subfigure[Formation switching from three-lane structure to two-lane structure]{
    \includegraphics[width=0.62\linewidth]{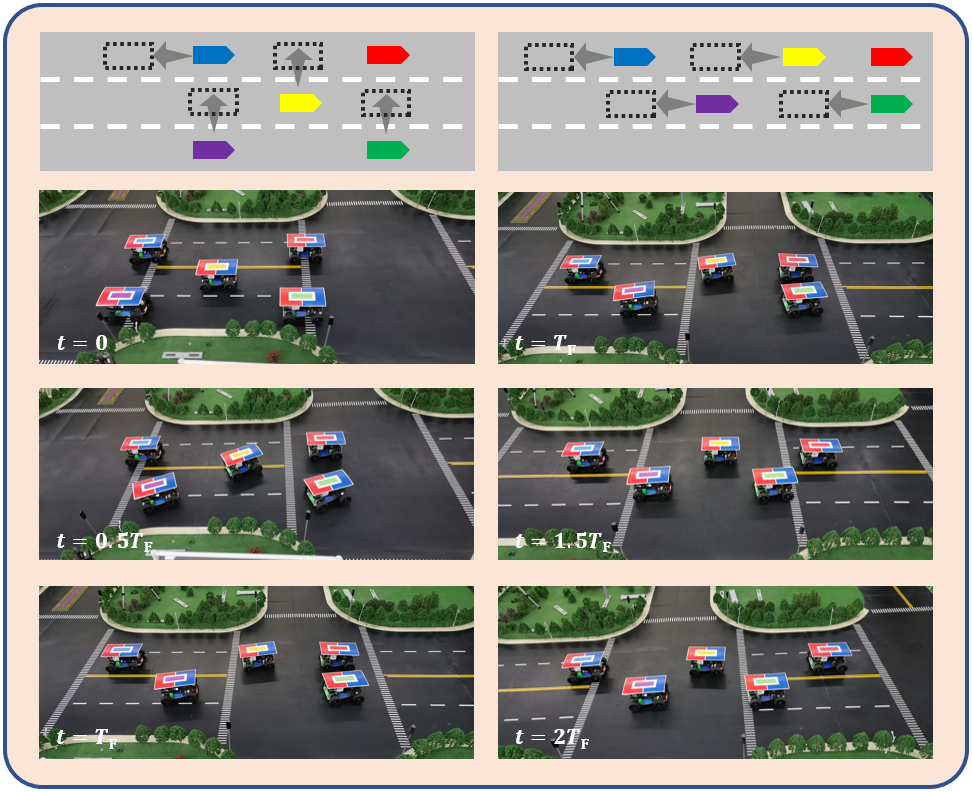}
    \label{formation1}}
    \subfigure[Formation switching from two-lane structure to one-lane structure]{
    \includegraphics[width=0.325\linewidth]{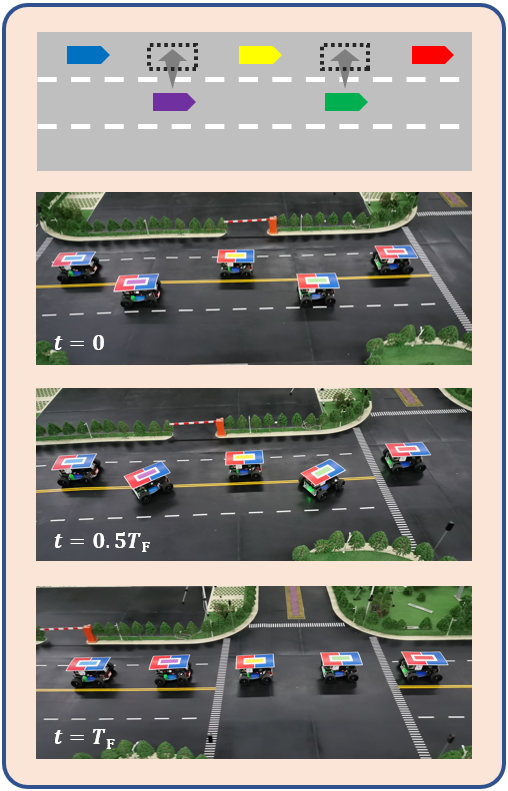}
    \label{formation1}}
    \caption{Snapshots of field experiments in lane number changing scenarios.}
    \label{snapshots1}
\end{center}
\end{figure*}

\begin{figure*}
\begin{center}
    \subfigure[Formation switching at on-ramp merging area]{
    \includegraphics[width=0.48\linewidth]{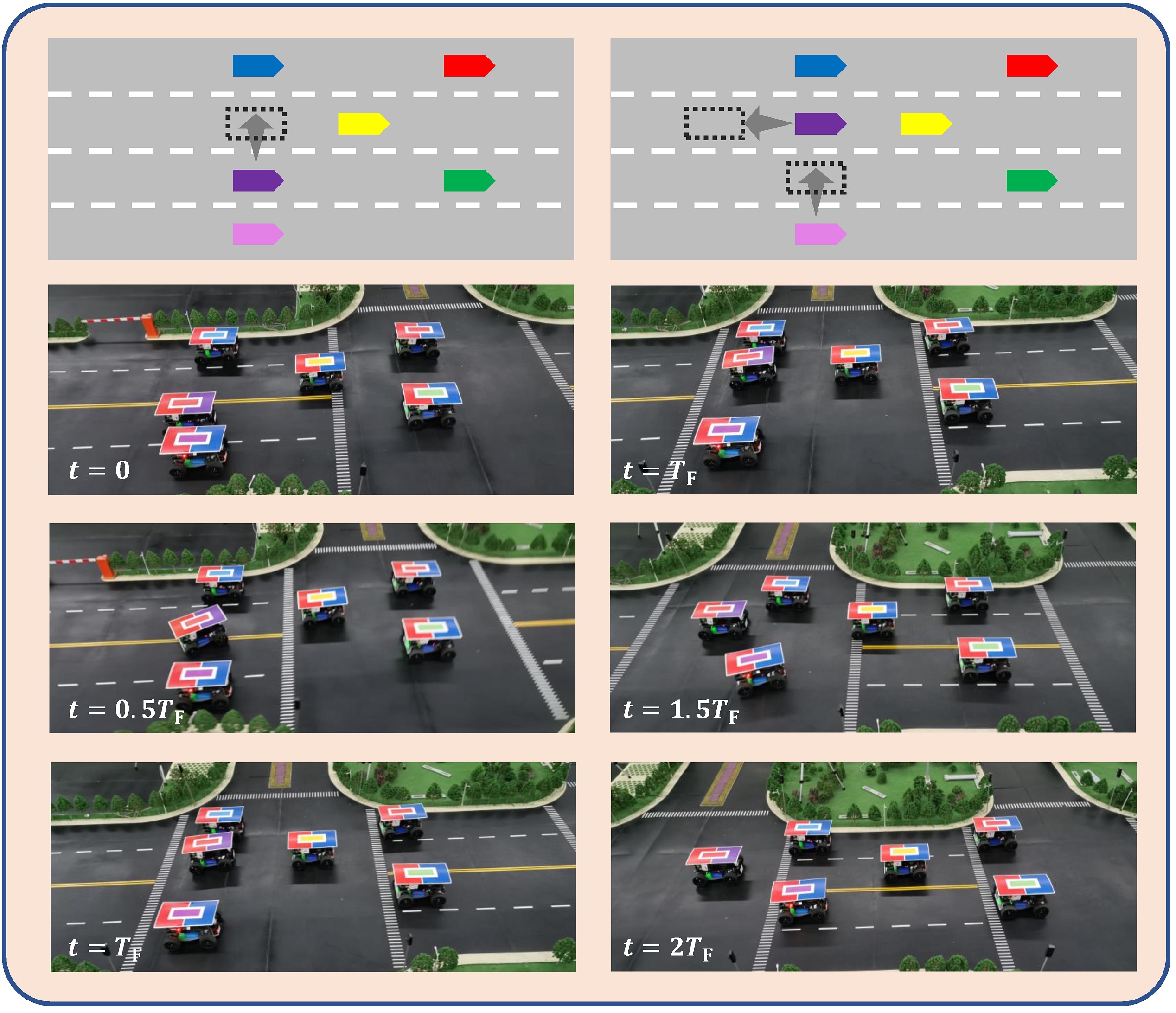}
    \label{formation1}}
    \subfigure[Formation switching at off-ramp leaving area]{
    \includegraphics[width=0.48\linewidth]{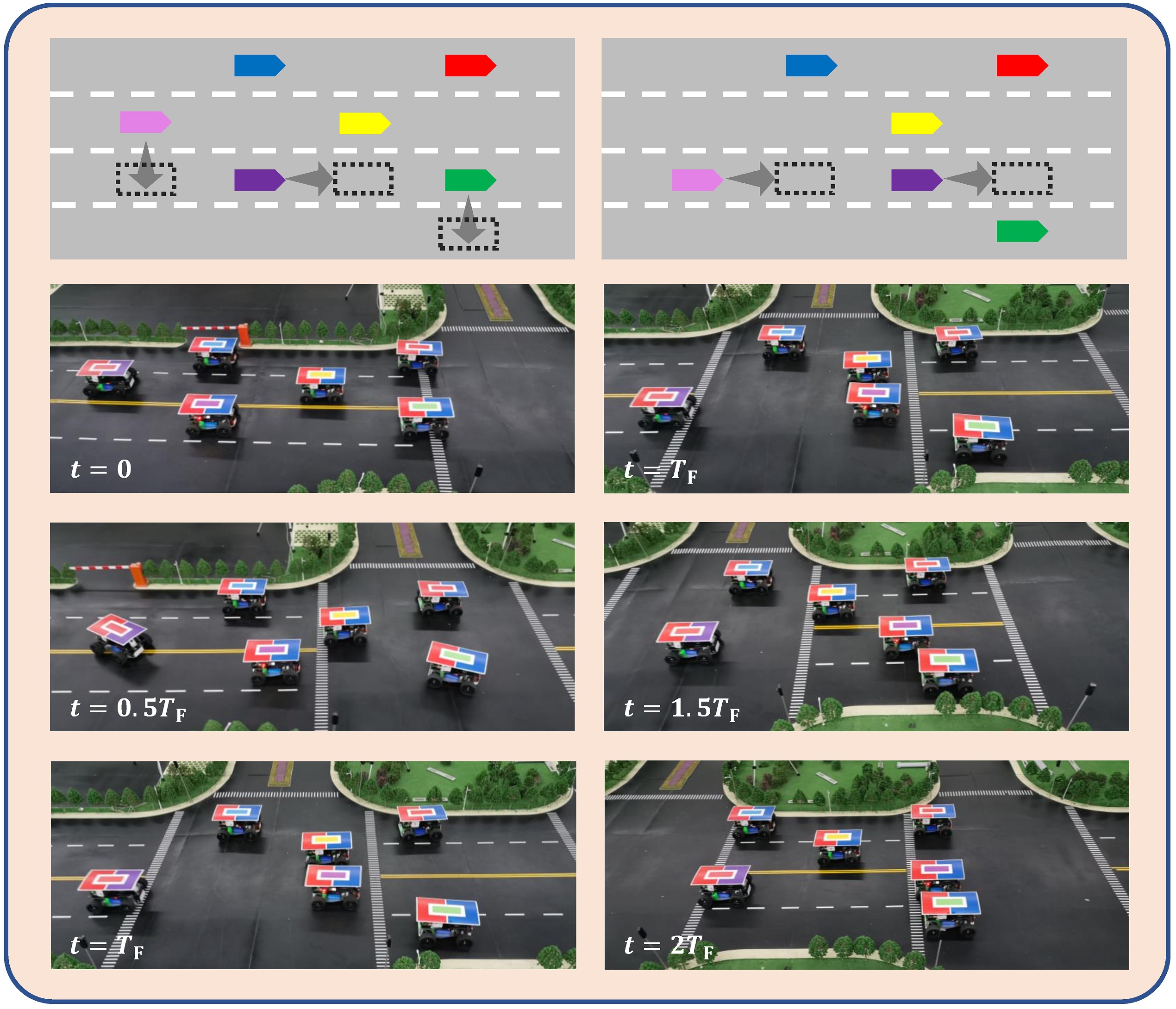}
    \label{formation1}}
    \subfigure[Formation switching in emergency leaving scenario]{
    \includegraphics[width=0.48\linewidth]{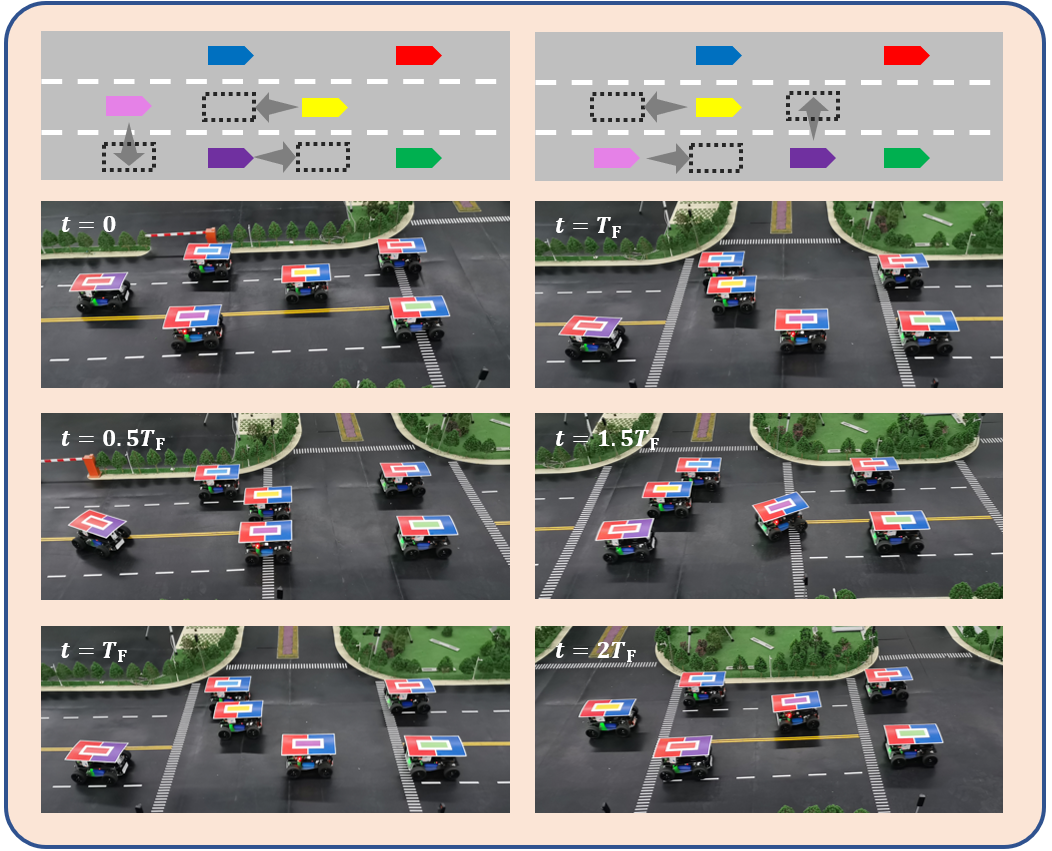}
    \label{formation1}}
    \subfigure[Formation switching in cooperative lane changing scenario]{
    \includegraphics[width=0.48\linewidth]{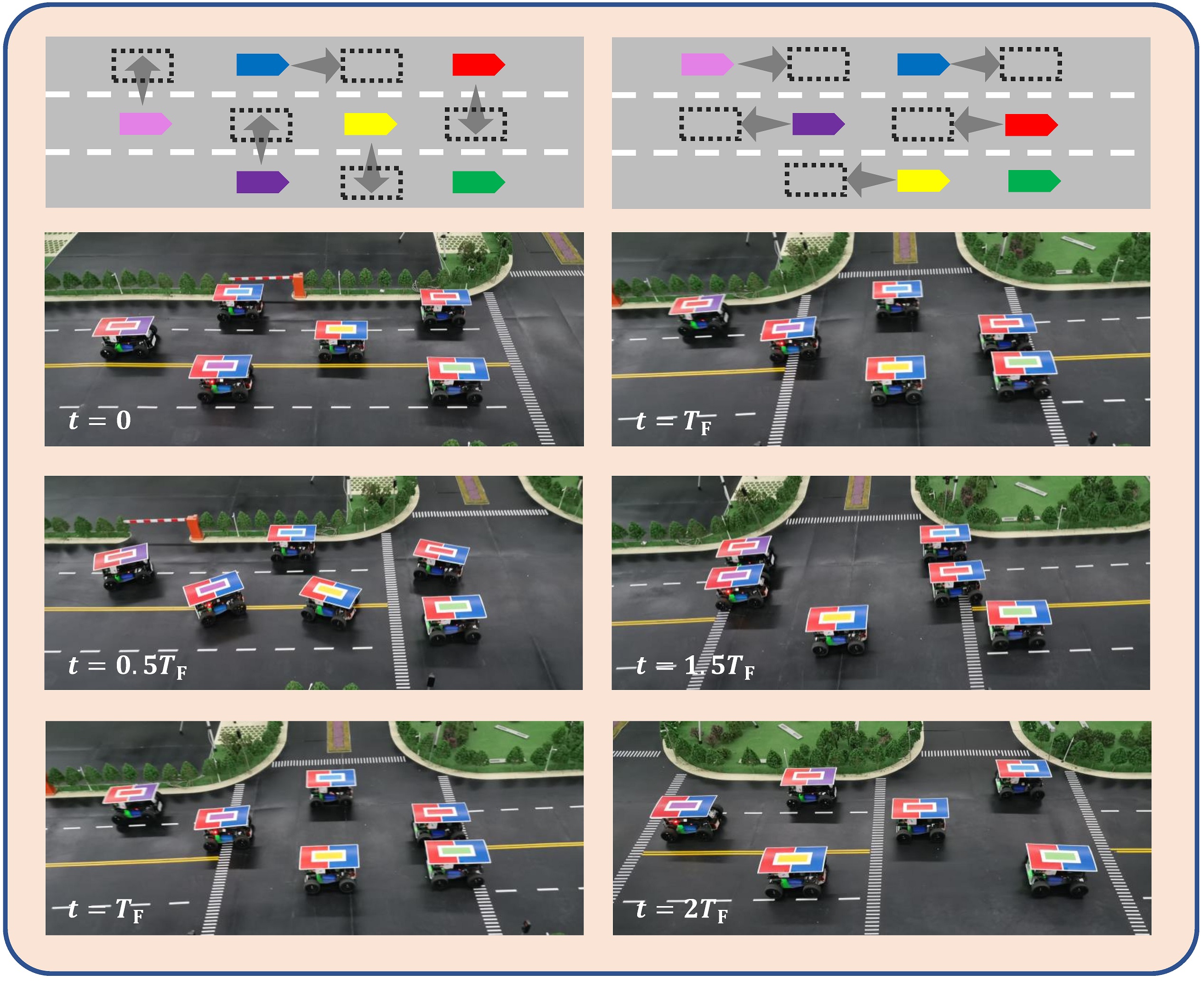}
    \label{formation1}}
    \caption{Snapshots of field experiments in formation structure switching scenarios.}
    \label{snapshots2}
\end{center}
\end{figure*}

As for the functional validation experiments, seven scenarios are designed, including lane number changing, ramp merging and leaving, and formation structure switching according to the demands of vehicles, as shown in Fig.~\ref{scenarios}. In scenario (a), the number of lanes change from one to three, and vehicles widen the formation to occupy all the drivable lanes and fully utilize the lane capacity. In scenario (b) and (c), the number of lanes decreases and vehicles have to narrow their structure to pass the bottleneck. In scenario (d) and (e), a vehicle attempts to join or leave a formation at the ramp area. In scenario (f) and (g), some vehicles are in emergency or have preference on specific lanes, so the formation has to switch its structure according to vehicles' demands. 

In scenarios (a), (b), (c), (e), (f), and (g), all the vehicles start from standard interlaced formation structure. In scenario (d), five vehicles start from a three-lane interlaced structure, and another vehicle starts on the rightmost lane. The desired speed of vehicles in all the scenarios are set as the same, which is $0.1\,\mathrm{m/s}$. The safe discretized distance $d_\text{F}$ is $0.5\,\mathrm{m}$. The steps that vehicles need to take and the snapshots of the experiments are presented in Fig.~\ref{snapshots1} and Fig.~\ref{snapshots2}. The videos of the process of experiments are available at the address: {\color{blue}https://github.com/cmc623/Formation-control-experiments}.

The presented snapshots and videos indicate that the proposed multi-lane formation control method is able to control multiple vehicles to form or switch to a desired geometric structure. In multiple scenarios, the centralized planner calculates the desired structure of formation and control inputs for vehicles, and vehicles switches the structure of formation step by step without collision, which validates the applicability of the proposed method in multiple scenarios.

Moreover, although this paper only carries out simulations and experiments in the multi-lane straight road scenario, it is apparently that the formation control method is able to be applied to more complex scenarios, \eg\,intersections and roundabouts. The lane preference in these complex scenarios can be transformed to target preference in formation, thus enables the application of multi-vehicle formation control.

%
\section{Conclusions}
\label{conc}
%

This paper introduces the multi-vehicle formation control method in multi-lane scenarios and carries out simulations and experiments to validate the performance of the method. The formation control framework is provided and the key methodologies are introduced in detail, including target generating, vehicle assignment, relative path planning, conflict resolution, trajectory planning, and tracking control of vehicles. Simulations are conducted with different number of vehicles and the performance of the utilized CBS method is compared with the priority-based A* method. Experiments are carried out in multiple scenarios on a micro-vehicle experimental platform. The results of simulations and experiments indicate that:
\begin{enumerate}
\item the formation control method, which utilizes CBS as the relative motion planning method, outperforms A* method in success rate, maximum steps and total steps of vehicles under different number of vehicles.
\item the formation control method is able to organize vehicles to switch structure of formations in multiple traffic scenarios, either with or without vehicles' preference on lanes.
\end{enumerate}

The future directions of this research include extending the proposed formation control framework to more complex scenarios and carry out more experiments to validate the performance.

\bibliographystyle{IEEEtranTIE}
\bibliography{thesis}

\begin{thebibliography}{10}
\providecommand{\url}[1]{#1}
\csname url@samestyle\endcsname
\providecommand{\newblock}{\relax}
\providecommand{\bibinfo}[2]{#2}
\providecommand{\BIBentrySTDinterwordspacing}{\spaceskip=0pt\relax}
\providecommand{\BIBentryALTinterwordstretchfactor}{4}
\providecommand{\BIBentryALTinterwordspacing}{\spaceskip=\fontdimen2\font plus
\BIBentryALTinterwordstretchfactor\fontdimen3\font minus
  \fontdimen4\font\relax}
\providecommand{\BIBforeignlanguage}[2]{{%
\expandafter\ifx\csname l@#1\endcsname\relax
\typeout{** WARNING: IEEEtran.bst: No hyphenation pattern has been}%
\typeout{** loaded for the language `#1'. Using the pattern for}%
\typeout{** the default language instead.}%
\else
\language=\csname l@#1\endcsname
\fi
#2}}
\providecommand{\BIBdecl}{\relax}
\BIBdecl

\bibitem{naus2010string}
G.~J. Naus, R.~P. Vugts, J.~Ploeg, M.~J. van De~Molengraft, and M.~Steinbuch,
  ``String-stable cacc design and experimental validation: A frequency-domain
  approach,'' \emph{IEEE Transactions on vehicular technology}, vol.~59, no.~9,
  pp. 4268--4279, 2010.

\bibitem{zheng2016distributed}
Y.~Zheng, S.~E. Li, K.~Li, F.~Borrelli, and J.~K. Hedrick, ``Distributed model
  predictive control for heterogeneous vehicle platoons under unidirectional
  topologies,'' \emph{IEEE Transactions on Control Systems Technology},
  vol.~25, no.~3, pp. 899--910, 2016.

\bibitem{wang2020controllability}
J.~Wang, Y.~Zheng, Q.~Xu, J.~Wang, and K.~Li, ``Controllability analysis and
  optimal control of mixed traffic flow with human-driven and autonomous
  vehicles,'' \emph{IEEE Transactions on Intelligent Transportation Systems},
  2020.

\bibitem{wu2020emergency}
J.~Wu, B.~Kulcs{\'a}r, S.~Ahn, and X.~Qu, ``Emergency vehicle lane
  pre-clearing: From microscopic cooperation to routing decision making,''
  \emph{Transportation research part B: methodological}, vol. 141, pp.
  223--239, 2020.

\bibitem{luo2016dynamic}
Y.~Luo, Y.~Xiang, K.~Cao, and K.~Li, ``A dynamic automated lane change maneuver
  based on vehicle-to-vehicle communication,'' \emph{Transportation Research
  Part C: Emerging Technologies}, vol.~62, pp. 87--102, 2016.

\bibitem{li2018consensus}
Y.~Li, C.~Tang, K.~Li, X.~He, S.~Peeta, and Y.~Wang, ``Consensus-based
  cooperative control for multi-platoon under the connected vehicles
  environment,'' \emph{IEEE Transactions on Intelligent Transportation
  Systems}, vol.~20, no.~6, pp. 2220--2229, 2018.

\bibitem{kato2002vehicle}
S.~Kato, S.~Tsugawa, K.~Tokuda, T.~Matsui, and H.~Fujii, ``Vehicle control
  algorithms for cooperative driving with automated vehicles and intervehicle
  communications,'' \emph{IEEE Transactions on intelligent transportation
  systems}, vol.~3, no.~3, pp. 155--161, 2002.

\bibitem{ntousakis2016optimal}
I.~A. Ntousakis, I.~K. Nikolos, and M.~Papageorgiou, ``Optimal vehicle
  trajectory planning in the context of cooperative merging on highways,''
  \emph{Transportation research part C: emerging technologies}, vol.~71, pp.
  464--488, 2016.

\bibitem{xu2020bi}
H.~Xu, Y.~Zhang, C.~G. Cassandras, L.~Li, and S.~Feng, ``A bi-level cooperative
  driving strategy allowing lane changes,'' \emph{Transportation research part
  C: emerging technologies}, vol. 120, p. 102773, 2020.

\bibitem{malikopoulos2018decentralized}
A.~A. Malikopoulos, C.~G. Cassandras, and Y.~J. Zhang, ``A decentralized
  energy-optimal control framework for connected automated vehicles at
  signal-free intersections,'' \emph{Automatica}, vol.~93, pp. 244--256, 2018.

\bibitem{bian2019cooperation}
Y.~Bian, S.~E. Li, W.~Ren, J.~Wang, K.~Li, and H.~X. Liu, ``Cooperation of
  multiple connected vehicles at unsignalized intersections: Distributed
  observation, optimization, and control,'' \emph{IEEE Transactions on
  Industrial Electronics}, vol.~67, no.~12, pp. 10\,744--10\,754, 2019.

\bibitem{yu2019corridor}
C.~Yu, Y.~Feng, H.~X. Liu, W.~Ma, and X.~Yang, ``Corridor level cooperative
  trajectory optimization with connected and automated vehicles,''
  \emph{Transportation Research Part C: Emerging Technologies}, vol. 105, pp.
  405--421, 2019.

\bibitem{cai2021formationb}
M.~Cai, Q.~Xu, C.~Chen, J.~Wang, K.~Li, J.~Wang, and Q.~Zhu, ``Formation
  control for connected and automated vehicles on multi-lane roads: Relative
  motion planning and conflict resolution,'' \emph{arXiv preprint
  arXiv:2103.10287}, 2021.

\bibitem{cai2021formationc}
M.~Cai, C.~Chen, J.~Wang, Q.~Xu, K.~Li, J.~Wang, and X.~Wu, ``Formation control
  with lane preference for connected and automated vehicles in multi-lane
  scenarios,'' \emph{arXiv preprint arXiv:2106.11763}, 2021.

\bibitem{chen2021mixed}
C.~Chen, J.~Wang, Q.~Xu, J.~Wang, and K.~Li, ``Mixed platoon control of
  automated and human-driven vehicles at a signalized intersection: dynamical
  analysis and optimal control,'' \emph{Transportation Research Part C:
  Emerging Technologies}, vol. 127, p. 103138, 2021.

\bibitem{dong2014time}
X.~Dong, B.~Yu, Z.~Shi, and Y.~Zhong, ``Time-varying formation control for
  unmanned aerial vehicles: Theories and applications,'' \emph{IEEE
  Transactions on Control Systems Technology}, vol.~23, no.~1, pp. 340--348,
  2014.

\bibitem{cheah2009region}
C.~C. Cheah, S.~P. Hou, and J.~J.~E. Slotine, ``Region-based shape control for
  a swarm of robots,'' \emph{Automatica}, vol.~45, no.~10, pp. 2406--2411,
  2009.

\bibitem{macdonald2011multi}
E.~A. Macdonald, ``Multi-robot assignment and formation control,'' Ph.D.
  dissertation, Georgia Institute of Technology, 2011.

\bibitem{li2016receding}
H.~Li, P.~Xie, and W.~Yan, ``Receding horizon formation tracking control of
  constrained underactuated autonomous underwater vehicles,'' \emph{IEEE
  Transactions on Industrial Electronics}, vol.~64, no.~6, pp. 5004--5013,
  2016.

\bibitem{marinescu2012ramp}
D.~Marinescu, J.~{\v{C}}urn, M.~Bouroche, and V.~Cahill, ``On-ramp traffic
  merging using cooperative intelligent vehicles: A slot-based approach,'' in
  \emph{2012 15th International IEEE Conference on Intelligent Transportation
  Systems}, pp. 900--906.\hskip 1em plus 0.5em minus 0.4em\relax IEEE, 2012.

\bibitem{marjovi2015distributed}
A.~Marjovi, M.~Vasic, J.~Lemaitre, and A.~Martinoli, ``Distributed graph-based
  convoy control for networked intelligent vehicles,'' in \emph{2015 IEEE
  Intelligent Vehicles Symposium (IV)}, pp. 138--143.\hskip 1em plus 0.5em
  minus 0.4em\relax IEEE, 2015.

\bibitem{navarro2016distributed}
I.~Navarro, F.~Zimmermann, M.~Vasic, and A.~Martinoli, ``Distributed
  graph-based control of convoys of heterogeneous vehicles using curvilinear
  road coordinates,'' in \emph{2016 IEEE 19th International Conference on
  Intelligent Transportation Systems (ITSC)}, pp. 879--886.\hskip 1em plus
  0.5em minus 0.4em\relax Ieee, 2016.

\bibitem{cai2019multi}
M.~Cai, Q.~Xu, K.~Li, and J.~Wang, ``Multi-lane formation assignment and
  control for connected vehicles,'' in \emph{2019 IEEE Intelligent Vehicles
  Symposium (IV)}, pp. 1968--1973.\hskip 1em plus 0.5em minus 0.4em\relax IEEE,
  2019.

\bibitem{xu2021coordinated}
Q.~Xu, M.~Cai, K.~Li, B.~Xu, J.~Wang, and X.~Wu, ``Coordinated formation
  control for intelligent and connected vehicles in multiple traffic
  scenarios,'' \emph{IET Intelligent Transport Systems}, vol.~15, no.~1, pp.
  159--173, 2021.

\bibitem{cai2021formationa}
M.~Cai, Q.~Xu, C.~Chen, J.~Wang, K.~Li, J.~Wang, and Q.~Zhu, ``Formation
  control for multiple connected and automated vehicles on multi-lane roads,''
  \emph{arXiv preprint arXiv:2103.10287}, 2021.

\bibitem{cai2021multi}
M.~Cai, Q.~Xu, C.~Chen, J.~Wang, K.~Li, J.~Wang, and X.~Wu, ``Multi-lane
  unsignalized intersection cooperation with flexible lane direction based on
  multi-vehicle formation control,'' \emph{arXiv preprint arXiv:2108.11112},
  2021.

\bibitem{zheng2021distance}
Y.~Zheng, Q.~Wang, D.~Cao, B.~Fidan, and C.~Sun, ``Distance-based formation
  control for multi-lane autonomous vehicle platoons,'' \emph{IET Control
  Theory \& Applications}, 2021.

\bibitem{cao2021platoon}
D.~Cao, J.~Wu, J.~Wu, B.~Kulcs{\'a}r, and X.~Qu, ``A platoon regulation
  algorithm to improve the traffic performance of highway work zones,''
  \emph{Computer-Aided Civil and Infrastructure Engineering}, 2021.

\bibitem{firoozi2021formation}
R.~Firoozi, X.~Zhang, and F.~Borrelli, ``Formation and reconfiguration of tight
  multi-lane platoons,'' \emph{Control Engineering Practice}, vol. 108, p.
  104714, 2021.

\bibitem{19kuhn1955hungarian}
H.~W. Kuhn, ``The hungarian method for the assignment problem,'' \emph{Naval
  research logistics quarterly}, vol.~2, no. 1-2, pp. 83--97, 1955.

\bibitem{hart1968formal}
P.~E. Hart, N.~J. Nilsson, and B.~Raphael, ``A formal basis for the heuristic
  determination of minimum cost paths,'' \emph{IEEE transactions on Systems
  Science and Cybernetics}, vol.~4, no.~2, pp. 100--107, 1968.

\bibitem{sharon2015conflict}
G.~Sharon, R.~Stern, A.~Felner, and N.~R. Sturtevant, ``Conflict-based search
  for optimal multi-agent pathfinding,'' \emph{Artificial Intelligence}, vol.
  219, pp. 40--66, 2015.

\bibitem{wagner2011m}
G.~Wagner and H.~Choset, ``M*: A complete multirobot path planning algorithm
  with performance bounds,'' in \emph{2011 IEEE/RSJ international conference on
  intelligent robots and systems}, pp. 3260--3267.\hskip 1em plus 0.5em minus
  0.4em\relax IEEE, 2011.

\bibitem{felner2018adding}
A.~Felner, J.~Li, E.~Boyarski, H.~Ma, L.~Cohen, T.~S. Kumar, and S.~Koenig,
  ``Adding heuristics to conflict-based search for multi-agent path finding,''
  in \emph{Proceedings of the International Conference on Automated Planning
  and Scheduling}, vol.~28, no.~1, 2018.

\bibitem{li2019improved}
J.~Li, A.~Felner, E.~Boyarski, H.~Ma, and S.~Koenig, ``Improved heuristics for
  multi-agent path finding with conflict-based search.'' in \emph{IJCAI}, vol.
  2019, pp. 442--449, 2019.

\bibitem{gonzalez2015review}
D.~Gonz{\'a}lez, J.~P{\'e}rez, V.~Milan{\'e}s, and F.~Nashashibi, ``A review of
  motion planning techniques for automated vehicles,'' \emph{IEEE Transactions
  on Intelligent Transportation Systems}, vol.~17, no.~4, pp. 1135--1145, 2015.

\bibitem{yang2021multi}
C.~Yang, J.~Dong, Q.~Xu, M.~Cai, H.~Qin, J.~Wang, and K.~Li, ``Multi-vehicle
  experiment platform: A digital twin realization method,'' \emph{arXiv
  preprint arXiv:2110.12859}, 2021.

\end{thebibliography}

\begin{IEEEbiography}[{\includegraphics[width=1in,height=1.25in,clip,keepaspectratio]{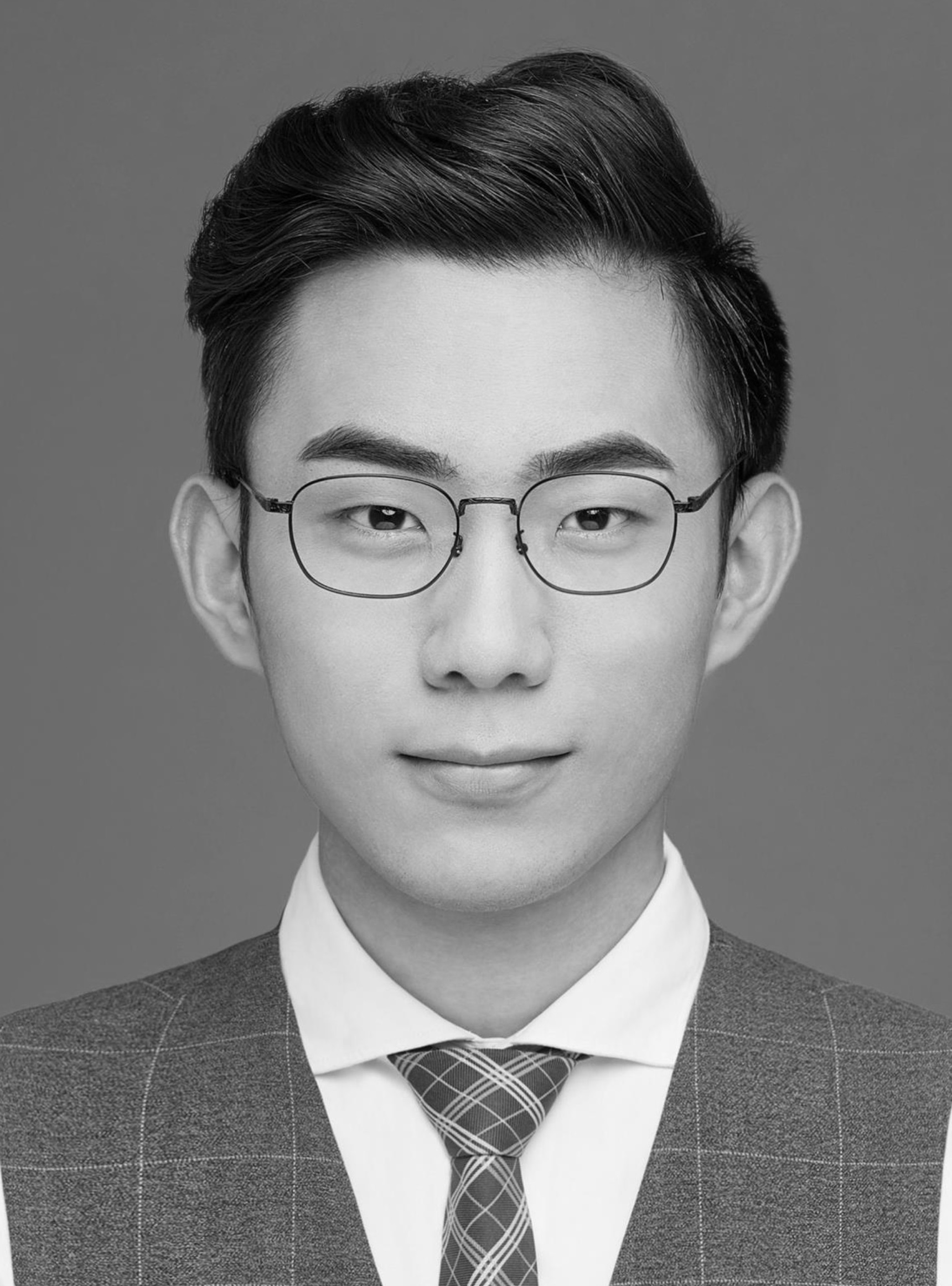}}]
	{Mengchi Cai} (Student Member, IEEE) received the B.E. degree from Tsinghua University, Beijing, China, in 2018. He is currently a Ph.D. candidate in mechanical engineering with the School of Vehicle and Mobility, Tsinghua University. His research interests include connected and automated vehicles, multi-vehicle formation control, and unsignalized intersection cooperation. 
\end{IEEEbiography}

\begin{IEEEbiography}[{\includegraphics[width=1in,height=1.25in,clip,keepaspectratio]{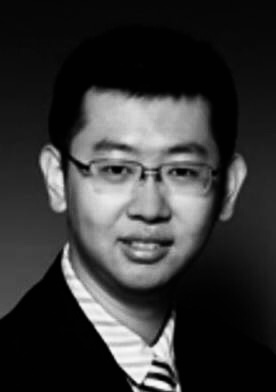}}]
	{Qing Xu} received his B.S. and M.S. degrees in automotive engineering from Beihang University, Beijing, China, in 2006 and 2008 respectively, and the Ph.D. degree in automotive engineering from Beihang University in 2014.
	
	During his Ph.D. research, he worked as a Visiting Scholar with the Department of Mechanical Science and Engineering, University of Illinois at Urbana–Champaign. From 2014 to 2016, he had his postdoctoral research in Tsinghua University. He is currently working as an Assistant Research Professor with the  School of Vehicle and Mobility, Tsinghua University. His main research interests include decision and control of intelligent vehicles.
\end{IEEEbiography}

\begin{IEEEbiography}[{\includegraphics[width=1in,height=1.25in,clip,keepaspectratio]{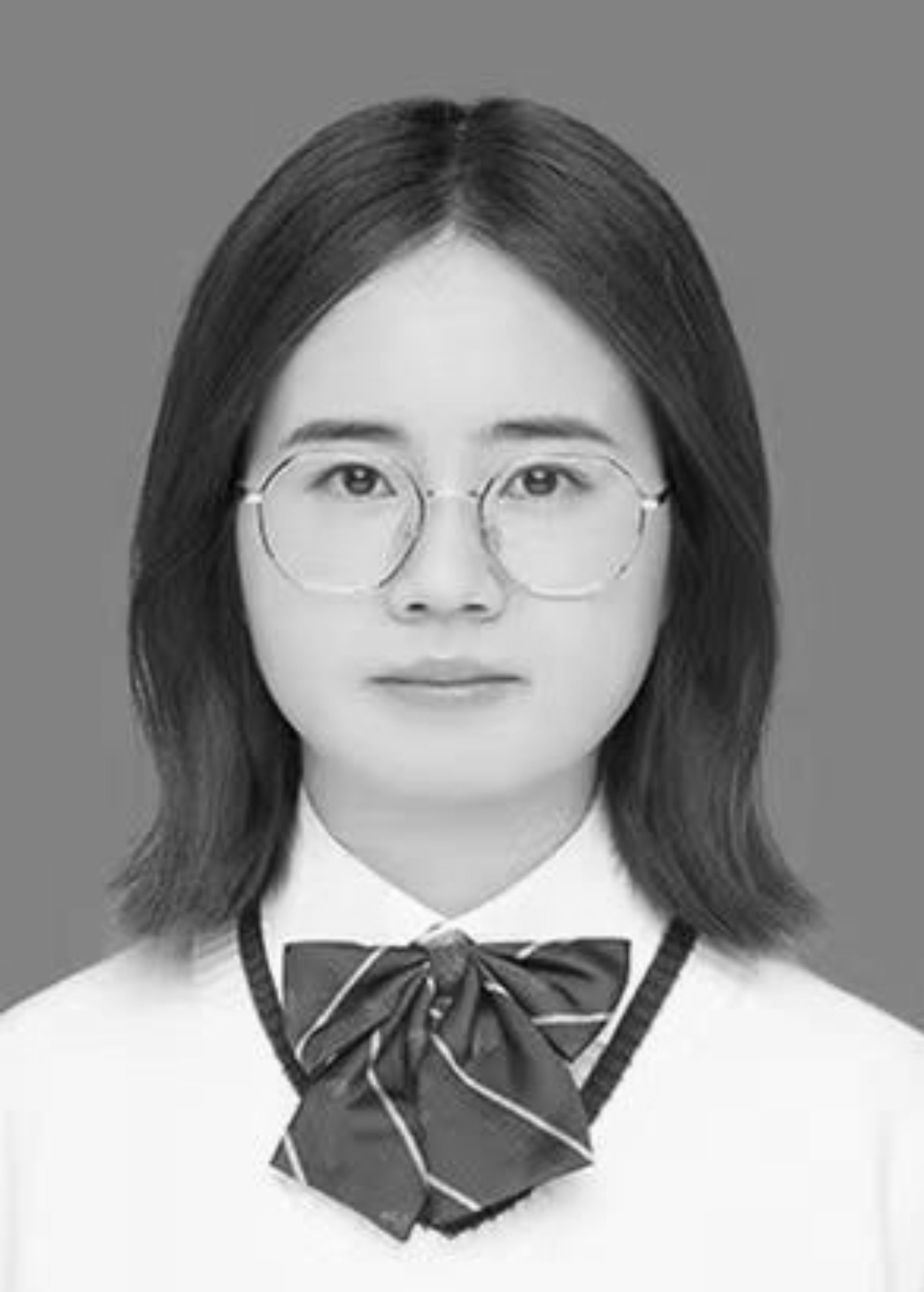}}]
	{Chunying Yang} received the B.E. degree from Beijingjiaotong University, Beijing, China, in 2019. She is currently a M.S. candidate in automotive engineering with the School of Transportation Science and Engineering, Beihang University. She is also a visiting student in School of Vehicle and Mobility, Tsinghua Univeristy. Her research interests include connected and automated vehicles, digital twin, and modeling of cloud control system.
\end{IEEEbiography}

\begin{IEEEbiography}[{\includegraphics[width=1in,height=1.25in,clip,keepaspectratio]{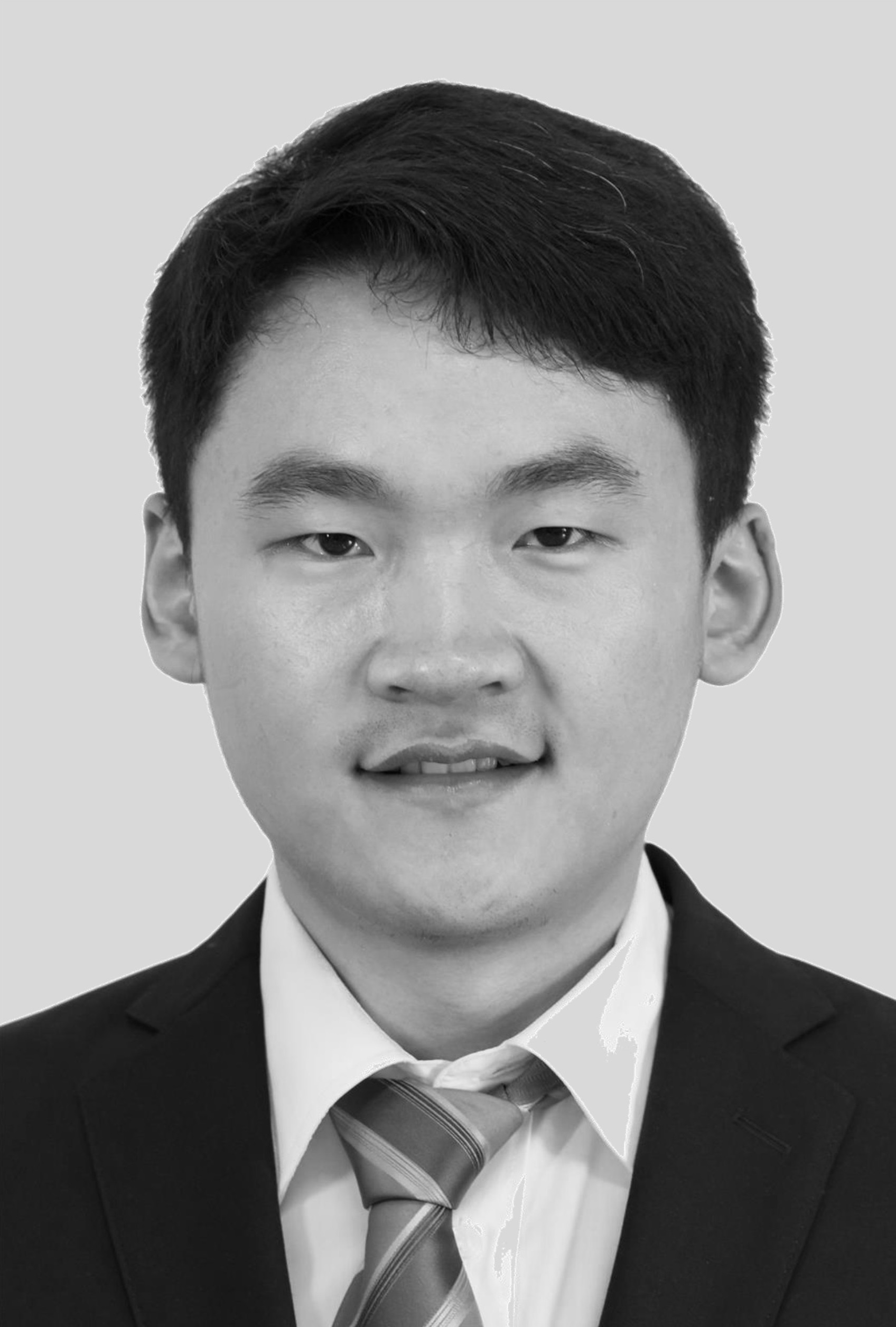}}]
	{Jianghong Dong} received the B.E. degree from Tsinghua University, Beijing, China, in 2020. He is currently a Ph.D. candidate in mechanical engineering with the School of Vehicle and Mobility, Tsinghua University. His research interests include connected and automated vehicles, multi-vehicle coordinated control, and vehicle-road-cloud integrated control.
\end{IEEEbiography}

\begin{IEEEbiography}[{\includegraphics[width=1in,height=1.25in,clip,keepaspectratio]{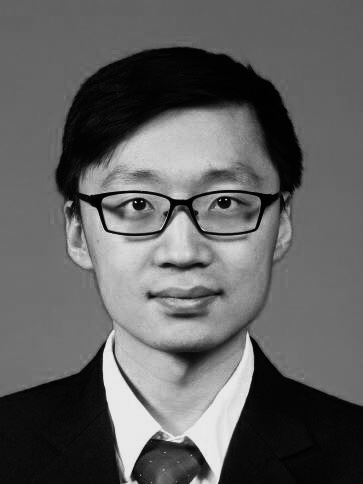}}]
	{Chaoyi Chen} (Student Member, IEEE) received the B.E. degree from Tsinghua University, Beijing, China, in 2016, and the M.S. from Tsinghua University, Beijing, China and RWTH Aachen University, Aachen, Germany in 2019. He is currently a Ph.D. student in mechanical engineering with the School of Vehicle and Mobility, Tsinghua University. He was a recipient of the Scholarship of Strategic Partnership RWTH Aachen University and Tsinghua University. His research interests include vehicular network, control theory and cooperative control.
\end{IEEEbiography}

\begin{IEEEbiography}[{\includegraphics[width=1in,height=1.25in,clip,keepaspectratio]{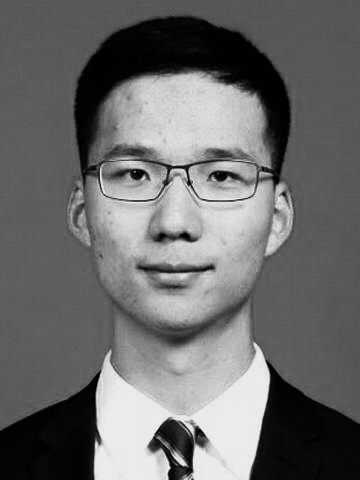}}]
{Jiawei Wang} (Student Member, IEEE) received the B.E. degree from Tsinghua University, Beijing, China, in 2018. He is currently a Ph.D. student in mechanical engineering with the School of Vehicle and Mobility, Tsinghua University. His research interests include connected automated vehicles, distributed control and optimization, and data-driven control. He was a recipient of the National Scholarship in Tsinghua University. He received the Best Paper Award at the 18th COTA International Conference of Transportation Professionals. 
\end{IEEEbiography}

\begin{IEEEbiography}[{\includegraphics[width=1in,height=1.25in,clip,keepaspectratio]{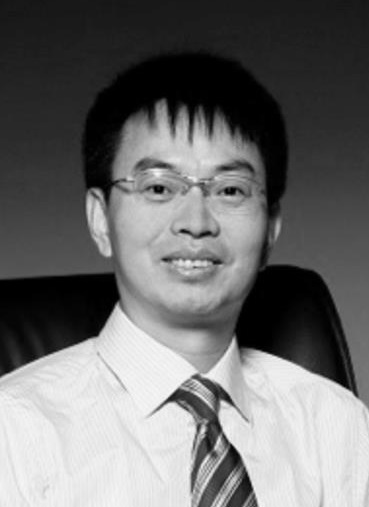}}]
{Jianqiang Wang} received the B. Tech. and M.S. degrees from Jilin University of Technology, Changchun, China, in 1994 and 1997, respectively, and the Ph.D. degree from Jilin University, Changchun, in 2002. He is currently a Professor with the School of Vehicle and Mobility, Tsinghua University, Beijing, China. 

He has authored over 150 papers and is a co-inventor of 99 patent applications. He was involved in over 10 sponsored projects. His active research interests include intelligent vehicles, driving assistance systems, and driver behavior. He was a recipient of the Best Paper Award in the 2014 IEEE Intelligent Vehicle Symposium, the Best Paper Award in the 14th ITS Asia Pacific Forum, the Best Paper Award in the 2017 IEEE Intelligent Vehicle Symposium, the Changjiang Scholar Program Professor in 2017, the Distinguished Young Scientists of NSF China in 2016, and the New Century Excellent Talents in 2008.
\end{IEEEbiography}

\begin{IEEEbiography}[{\includegraphics[width=1in,height=1.25in,clip,keepaspectratio]{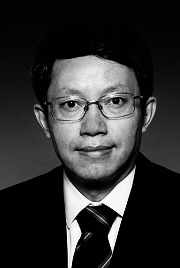}}]{Keqiang Li} received the B.Tech. degree from Tsinghua University of China, Beijing, China, in 1985, and the M.S. and Ph.D. degrees in mechanical engineering from the Chongqing University of China, Chongqing, China, in 1988 and 1995, respectively.
	
	He is currently a Professor with the School of Vehicle and Mobility, Tsinghua University. His main research areas include automotive control system, driver assistance system, and networked dynamics and control. He is leading the national key project on CAVs (Intelligent and Connected Vehicles) in China. Dr. Li has authored more than 200 papers and is a co-inventor of over 80 patents in China and Japan.
	
	Dr. Li has served as Fellow Member of Society of Automotive Engineers of China, editorial boards of the \emph{International Journal of Vehicle Autonomous Systems}, Chairperson of Expert Committee of the China Industrial Technology Innovation Strategic Alliance for CAVs (CACAV), and CTO of China CAV Research Institute Company Ltd. (CCAV). He has been a recipient of Changjiang Scholar Program Professor, National Award for Technological Invention in China, etc.
\end{IEEEbiography}

\balance

\end{document}